\documentclass[journal]{IEEEtran}

\usepackage{graphicx}
\usepackage{color}
\usepackage{mathrsfs}
\usepackage{amsmath, nicefrac}
\usepackage{multirow}
\usepackage{amsfonts}
\usepackage[switch]{lineno}
\usepackage{balance}
\usepackage{subfigure}
\usepackage{tikz}
\usepackage{tikz-3dplot}
\usepackage{pgfplots}

\usepackage{booktabs}

\usepackage{enumerate} 

\usepackage{algorithm}
\usepackage{algpseudocode}

\DeclareMathOperator*{\argmax}{arg\,max}

\usetikzlibrary{positioning,calc,arrows,shadows,decorations.pathreplacing}
\usetikzlibrary{external}

\definecolor{myyellow}{RGB}{255,250,225}
\definecolor{myred}{RGB}{255,206,206}
\definecolor{myblue}{RGB}{231,239,246}
\definecolor{mygreen}{RGB}{201,223,138}
\definecolor{new_gray}{RGB}{164,188,196}
\definecolor{new_yell}{RGB}{254,246,235}
\definecolor{new_blue}{RGB}{217,224,226}
\definecolor{new_gree}{RGB}{244,243,243}
\definecolor{new_beig}{RGB}{190,185,181}
\definecolor{new_gree_2}{RGB}{238,246,236}

\begin{document}

\title{Discriminative Singular Spectrum Classifier with Applications on Bioacoustic Signal Recognition}

\author{Bernardo B. Gatto, Juan G. Colonna, Eulanda M. dos Santos, Alessandro L. Koerich and Kazuhiro Fukui
\thanks{B. B. Gatto is with Center for Artificial Intelligence Research (C-AIR),  Tsukuba, Japan e-mail: bernardo@cvlab.cs.tsukuba.ac.jp}%
\thanks{J. G. Colonna is with Institute of Computing, Federal University of Amazonas, Manaus, AM, Brazil e-mail: juancolonna@icomp.ufam.edu.br}%
\thanks{E. M. dos Santos is with Institute of Computing, Federal University of Amazonas, Manaus, AM, Brazil e-mail: emsantos@icomp.ufam.edu.br}%
\thanks{A. L. Koerich is with \'Ecole de Technologie Sup\'erieure (\'ETS), Universit\'e du Qu\'ebec, Montreal, QC, Canada e-mail: alessandro.koerich@etsmtl.ca}
\thanks{K. Fukui is with Center for Artificial Intelligence Research (C-AIR),  Tsukuba, Japan e-mail: kfukui@cs.tsukuba.ac.jp}%
\thanks{Manuscript submitted. March, 2021}
}

\markboth{Submitted to IEEE/ACM Transactions on Audio Speech and Language Processing.}
{Gatto \MakeLowercase{\textit{et al.}}: Discriminative Singular Spectrum Classifier with Applications on Bioacoustic Signal Recognition}

\maketitle
\begin{abstract}

Automatic analysis of bioacoustic signals is a fundamental tool to evaluate the vitality of our planet. Frogs and bees, for instance, may act like biological sensors providing information about environmental changes. This task is fundamental for ecological monitoring and includes many challenges such as nonuniform signal length processing, degraded target signal due to environmental noise, and the scarcity of the labeled samples for training machine learning. To tackle these challenges, we present a bioacoustic signal classifier equipped with a discriminative mechanism to extract useful features for analysis and classification efficiently. The proposed classifier does not require a large amount of training data and handles nonuniform signal length natively. Unlike current bioacoustic recognition methods, which are task-oriented, the proposed model relies on transforming the input signals into vector subspaces generated by applying Singular Spectrum Analysis (SSA). Then, a subspace is designed to expose discriminative features. The proposed model shares end-to-end capabilities, which is desirable in modern machine learning systems. This formulation provides a segmentation-free and noise-tolerant approach to represent and classify bioacoustic signals and a highly compact signal descriptor inherited from SSA. The validity of the proposed method is verified using three challenging bioacoustic datasets containing anuran, bee, and mosquito species. Experimental results on three bioacoustic datasets have shown the competitive performance of the proposed method compared to commonly employed methods for bioacoustics signal classification in terms of accuracy.

\end{abstract}

\begin{IEEEkeywords}
Bioacoustic Signal Classification, Singular Spectrum Analysis, Mutual Singular Spectrum Analysis, Signal Subspace Methods.
\end{IEEEkeywords}
\begin{center} \bfseries EDICS Category: AUD-CLAS \end{center}
%
\IEEEpeerreviewmaketitle

\section{Introduction}
%
%
%

\IEEEPARstart{E}{nvironmental} monitoring has been taking an increasingly important role by providing the means to analyze and evaluate climate changes. Tasks as cataloging and counting animals through bioacoustic monitoring provide a large amount of information that can generate knowledge to understand and solve diverse problems. For instance, recent studies have pointed out that some species of birds' migratory route has been drastically affected by global warming~\cite{global_1,global_2,global_3,global_4}. Since these animals are sensitive to such changes, it is valuable to study their populations' dynamics over time.

Ecological monitoring has many challenges, such as obtaining information from remote access areas and the use of specialized equipment, which are often expensive. For invertebrate species, for instance, population monitoring is usually based on using traps to measure the population density at a given location~\cite{trap_1,trap_2}. However, the use of a large number of traps is problematic because it can be expensive and harmful to agricultural landscapes that depend on pollinating insects. The use of traps implies that the task involved in counting the individuals is performed manually, increasing the monitoring cost. Besides, it may cause an ecological imbalance since the frequent use of traps may directly interfere with the ecosystem where a particular species can live~\cite{unb_1,unb_2}.
To cope with the challenges mentioned above, several authors have presented solutions based on bioacoustic signal classification. Solutions based on passive acoustic recorders have minimal impact on the ecosystem and can be implemented with low-cost hardware~\cite{low_1,low_2}. These solutions are usually integrated with a sensor network to capture signals in scattered geographic locations. Such signals can be processed locally in devices attached to the sensors or can be sent through the network nodes. During this processing stage, a classification model may be employed to count individuals and send just these results, decreasing the network's data load.

\begin{figure}[htpb!]
\centering
\includegraphics[width=\linewidth]{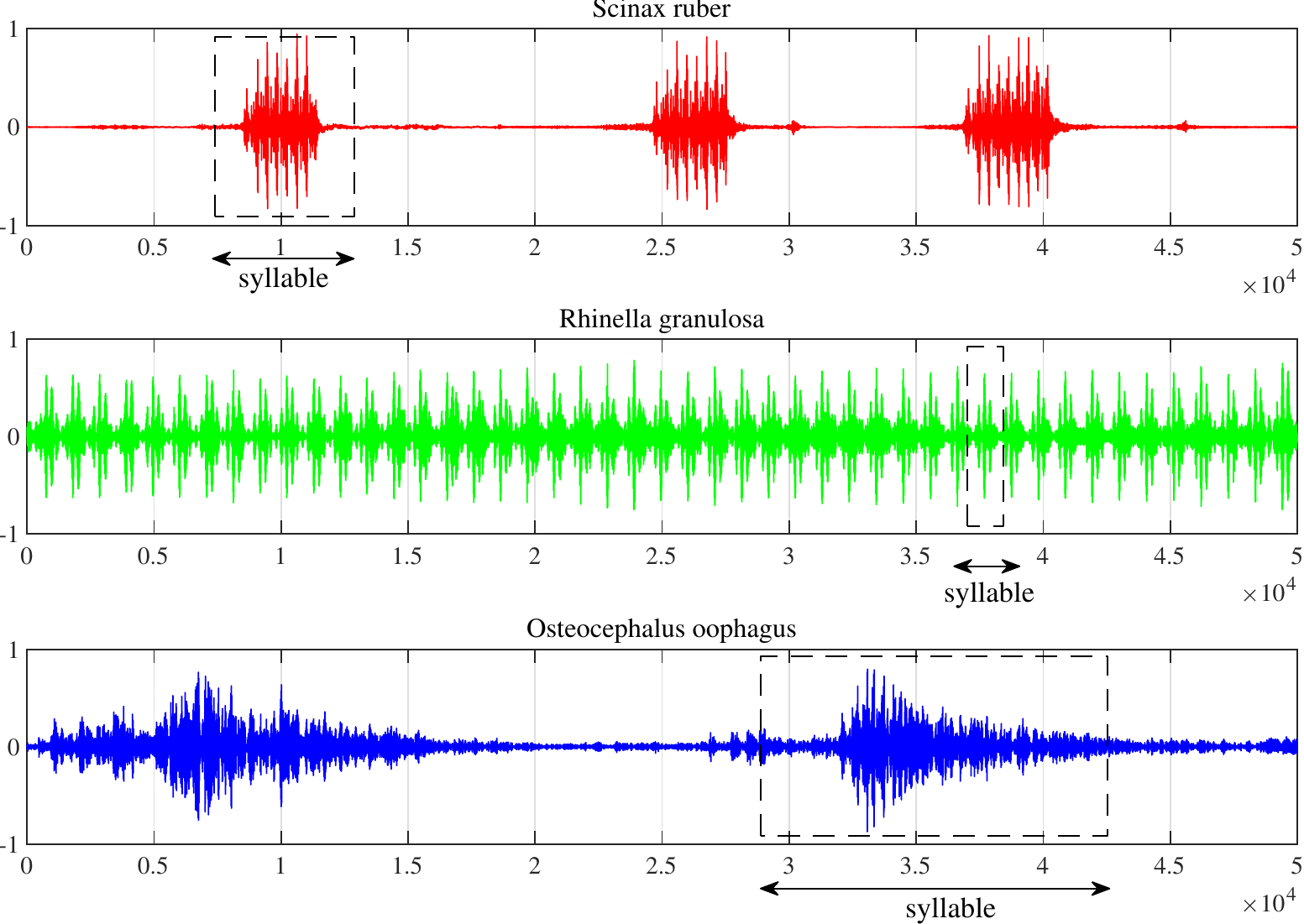}
\caption{Challenges of comparing bioacoustic signals due to the variation of syllable lengths according to the anuran species, and the signal synchronization. Besides, substantial parts of the signals are not informative (e.g., the segment between $0$ and $0.5$ of the Scinax ruber call).}
\label{fig:different_signals}
\end{figure}

Classical methods for bioacoustic signal recognition are task-oriented systems because they separate the main task into four fundamental steps: environmental noise removal, syllable segmentation, feature engineering, and classification. Despite their performance, these methods cannot be embedded in low-cost hardware due to the computational complexity and the memory requirements of each step. For example, methods based on syllable segmentation generally employ iterative algorithms, which are time-consuming. Besides, most machine learning approaches require input signals of fixed size, which makes deployment difficult since the syllables of bioacoustic signals can be arbitrary in length~\cite{len_1,len_2}.
Fig.~\ref{fig:different_signals} shows some challenges in recognizing bioacoustic signals. The recordings of three anuran species -- \textit{Scinax ruber}, \textit{Rhinella granulosa}, and \textit{Osteocephalus oophagus} -- vary in syllable length and have different alignments, requiring sophisticated segmentation methods and robust feature extraction techniques~\cite{Xie2020}. Since these syllables have a variable length, it is difficult to adjust a single fixed-length temporal window to segment these signals. A short-term window can result in excessive fragmentation of these syllables, while a long-term window ends up including long segments of environmental noise or stretches of contiguous syllables. Overall, the recordings may also present a high level of redundancy and long segments with no informative data, i.e., segments with only ambient background sound. 
Autonomous bioacoustic monitoring systems may exhibit additional requirements which current approaches may not fulfill. For example, the classification models should be available in a lightweight computational design, allowing its implementation on resource-constrained hardware~\cite{low_1,low_2}. Another requirement is related to efficient models generated with few training samples. Labeled data of some species may be scarce or may not even be accessible (e.g., data from endangered species) due to the difficulty of recording them, demanding training under small-scale datasets~\cite{trap_1,trap_2}.

Recently, new bioacoustics methods have emerged based on the subspace analysis theory of the autocorrelation matrix to circumvent the issues mentioned above~\cite{recent_1,recent_2}.
Subspace-based methods group signals into clusters called subspaces. These subspaces are defined in a high-dimensional vector space, where the learning patterns are represented as a linear combination of several basis vectors. Such basis vectors are ranked according to their information contribution retained by the eigenvalues, providing a data compression and selection mechanism. Since acoustic sensors may collect signals with information overlap, these methods can compress those signals, proving a compact representation through a subset of their eigenvectors. In general, subspace-based methods operate on multiple patterns at once, achieving higher recognition rates than methods that operate on single patterns~\cite{subspace_1,subspace_2,subspace_3}.

Among the subspace-based methods, Mutual Singular Spectrum Analysis (MSSA) is designed to handle signals of nonuniform length, achieving competitive results on supervised learning problems~\cite{mssa,gssa}. MSSA, also called Singular Spectrum Classifier, employs basis vectors obtained by Singular Spectrum Analysis (SSA) to represent bioacoustic signals. As basis vectors span a subspace, the comparison among bioacoustic signals is simplified by the use of canonical angles. This method achieved encouraging results in very challenging datasets~\cite{mssa,gssa}. Moreover, MSSA is computationally efficient. It requires only one singular value decomposition (SVD) transform to represent a bioacoustic signal of any length. It allows embedding a classification model in a device attached to an acoustic sensor with limited hardware resources. Benefits of employing MSSA include its capacity of handling signals of any length and its high compression capability. Another advantage of MSSA is its relative robustness to environmental noise with Gaussian characteristics, such as additive white Gaussian noise (AWGN) or additive colored Gaussian noise (ACGN)~\cite{g_noise_1,g_noise_2}. Besides, MSSA includes an automatic feature extraction mechanism, which does not require the use of any external feature extraction technique. Since the method operates directly on signals of different lengths, the extraction of syllables becomes unnecessary, making the method computationally efficient. As a result, MSSA is a time and memory-efficient method, which is highly desirable in bioacoustic signal classification applications.

Employing subspaces presents the benefit of appropriately representing signals even when few learning samples are available. The linear subspaces express data through the linear combination of features. Thus, basis vectors extracted from a few examples may represent many signals, since the linear correlation between them is commonly high. This advantage allows subspace-based methods to achieve excellent results even when few learning samples are available.
However, despite its computational efficiency and benefits, MSSA has no discriminative mechanism. The subspaces generated to represent the bioacoustic signals may not be optimal for a classification task, since they are computed independently, neglecting the relationship that may exist between subspace generation and class discrimination. This drawback may prevent MSSA from achieving even more competitive results.

\begin{figure}
	\centering
	\scalebox{.85}{%
	
	\tikzset{
		every node/.style={text centered,draw,minimum height=0.7cm, minimum width=2.8cm, node distance=20pt}
	}
	\begin{tikzpicture}[font=\normalsize]
	\node  [coordinate]  at (0,0.5)  (A){A};
	\node  [right=of A, yshift=-0.5cm,font=\normalsize,fill=new_gree_2] (C)
    {input signal};
    
    \node  [below=of C, xshift=1.75cm, yshift=0.5cm, font=\normalsize,fill=gray!10] (D)
    {noise filtering};
    
	\node  [right=of C,   font=\normalsize,fill=gray!10] (CP1){signal segm.};
	
	\node  [below=of CP1, xshift=1.75cm, yshift=0.5cm, font=\normalsize,fill=gray!10] (E)
    {feature sel.};
	
	\node  [right=of CP1, font=\normalsize,fill=gray!20] (CP2) {classification};


	\draw [->, >=latex] (C) -| (D);
	\draw [->, >=latex] (D) -| (CP1);
	\draw [->, >=latex] (CP1) -| (E);
	\draw [->, >=latex] (E) -| (CP2);
	
	\end{tikzpicture}
	}%
\caption{Conceptual figure of a task-oriented system.}
\label{fig:conventional_pipeline}
\end{figure}
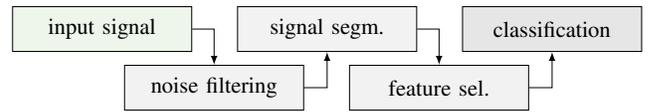
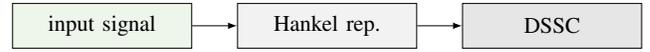
\begin{figure}
	\centering
	\scalebox{.85}{%
	
	\tikzset{
		every node/.style={text centered,draw,minimum height=0.7cm, minimum width=2.8cm, node distance=20pt}
	}
	\begin{tikzpicture}[font=\normalsize]
	\node  [coordinate]  at (0,0.5)  (A){A};
	\node  [right=of A, yshift=-0.5cm,font=\normalsize,fill=new_gree_2] (C)
    {input signal};
	\node  [right=of C,   font=\normalsize,fill=gray!10] (CP1){Hankel rep.};
	\node  [right=of CP1, font=\normalsize,fill=gray!20] (CP2) {DSSC};
	\draw [->, >=latex] (C) -- (CP1);
	\draw [->, >=latex] (CP1) -- (CP2);
	\end{tikzpicture}
	}%
	\caption{Conceptual figure of the proposed system.}
	\label{fig:framework}
\end{figure}

	
	
	

In light of these facts and motivated by the recent results achieved by MSSA~\cite{mssa,gssa}, in this paper, we propose a discriminative method for bioacoustic recognition called Discriminative Singular Spectrum Classifier (DSSC) as an extension of MSSA. 
DSSC is designed by incorporating a mechanism of extracting discriminative features based on the projection onto the generalized difference subspace (GDS)~\cite{diff} into the framework of MSSA. The essence of DSSC is to conduct MSSA on a GDS, which is calculated from the training subspaces generated by SSA. More concretely, in DSSC, all subspaces are projected on a GDS before measuring the canonical angles between them. Since a GDS contains mainly the components' difference among the reference subspaces, GDS projection can enlarge the angles between them toward the orthogonal status. Thus, DSSC presents a high discriminative ability. It classifies the subspaces projected on GDS that inherit discriminative features extracted from the training subspaces.

The effectiveness of the GDS projection has been demonstrated in several image recognition tasks such as face and 3D object recognition \cite{diff}. However, to the best of our knowledge, our DSSC is the first trial in which the GDS projection is applied to a task of signal classification, in particular, focusing on bioacoustic signal classification. DSSC inherits the advantages of MSSA, such as the compact subspace representation, the ability to handle signals of different lengths without segmentation, robustness to noise, and high capacity to learn from small-scale training sets, showing higher discriminative power compared to MSSA. Fig.~\ref{fig:conventional_pipeline} presents the conventional pipeline, where the input signal goes through a pipeline containing noise filtering, signal segmentation, feature selection, and classification. On the other hand, Fig.~\ref{fig:framework} shows the proposed framework, which presents fewer learnable modules, providing a lightweight system.

In addition to its advantages, DSSC shares some capacities observed in End-to-End (E2E) systems, where a single model learns to solve a complex task that describes the entire target system. E2E systems usually avoid the intermediate layers existing in traditional pipelines.
DSSC bypasses some restrictions of E2E systems, such as the demand for a massive amount of training data and the difficulty to improve or modify the system (e.g., increasing or decreasing the target species in the sensor node).

The main contributions of this paper are summarized as follows:
\begin{enumerate}[(i)]
\item We demonstrate that GDS projection can enhance the discrimination for signal subspaces generated through SSA.
\item We propose a more discriminative method based on subspace representation, which is called DCCA, for bioacoustic signal classification. This method equips a powerful feature extraction by GDS projection as a powerful extension of MSSA.
\item We verify the effectiveness of DCCA for signal classification through extensive evaluations on various types of bioacoustic signal datasets.
\end{enumerate}

This paper is organized as follows. In Section~\ref{sec:related}, we present a brief review on bioacoustic signal classification. In Section~\ref{sec:proposed}, we describe the proposed method, as well as its application on bioacoustic signal classification. Experimental results are presented in Section~\ref{sec:experimental}. Finally, conclusion and future work are discussed in the last section. 

\section{Related Work}
\label{sec:related}
\begin{figure}[t!]
	\centering
	\scalebox{.675}{%
	\begin{tikzpicture}

    \fill[gray!20] (2.6,1) rectangle (4,5);
    \node[] at (0,2.8) {\Large ${X}$};
    \draw[<->] (2.5,0.8) -- (4.1,0.8) node[anchor=north west] {$l$};
    \draw[->] (6.4, 2.6) -- (7.4, 2.6) node[anchor=north west] {};

\begin{axis}[
    axis lines=none
]

\addplot[
    color=blue,
    ]
    coordinates { (1	,0.000274658)(2	,-0.000762939)(3	,-0.000213623)(4	,0.000549316)(5	,-0.000305176)(6	,-0.000976563)(7	,-0.000183105)(8	,-0.000152588)(9	,-0.000762939)(10	,-0.000701904)(11	,-0.001739502)(12	,0.000274658)(13	,0.000366211)(14	,0.001220703)(15	,-0.001922607)(16	,-0.000488281)(17	,0.000427246)(18	,0.001312256)(19	,-0.001159668)(20	,-0.000671387)(21	,-0.000427246)(22	,-0.001037598)(23	,-0.000640869)(24	,-0.000732422)(25	,0.000305176)(26	,-0.002258301)(27	,-0.001251221)(28	,-0.002227783)(29	,0.001953125)(30	,0.000579834)(31	,0.000366211)(32	,-0.001556396)(33	,0.000274658)(34	,0.002655029)(35	,0.00213623)(36	,0.000732422)(37	,-0.00088501)(38	,0.001342773)(39	,0.002349854)(40	,0.002044678)(41	,-0.000427246)(42	,0.000396729)(43	,0.000366211)(44	,-0.000518799)(45	,-0.00112915)(46	,-0.00112915)(47	,-0.000488281)(48	,-0.000518799)(49	,0)(50	,0.000213623)(51	,0.000701904)(52	,0.000518799)(53	,-0.000762939)(54	,-0.000762939)(55	,0.002258301)(56	,-0.000762939)(57	,-0.000183105)(58	,-0.00012207)(59	,0.002746582)(60	,-0.000305176)(61	,-0.000549316)(62	,-0.002197266)(63	,0.000762939)(64	,0.000854492)(65	,0.001922607)(66	,0.000366211)(67	,0.000274658)(68	,-0.000396729)(69	,0)(70	,0.000549316)(71	,0.000762939)(72	,-0.000854492)(73	,-0.001525879)(74	,0)(75	,0.000244141)(76	,-0.002319336)(77	,-0.000854492)(78	,-0.002502441)(79	,-0.000488281)(80	,0)(81	,-0.000793457)(82	,-0.000457764)(83	,-0.00012207)(84	,0.000335693)(85	,0.00112915)(86	,-0.000274658)(87	,-0.000579834)(88	,0.000610352)(89	,0.000152588)(90	,0.001312256)(91	,-0.001922607)(92	,0.001800537)(93	,-0.001525879)(94	,0.004821777)(95	,-0.004058838)(96	,0.003753662)(97	,-0.003173828)(98	,0.001647949)(99	,-0.001281738)(100	,-0.001037598)(101	,0.002105713)(102	,-0.002929688)
 
    };
\end{axis}

 	\node[] at (10.1, 2.8) {$\begin{bmatrix}
 			x_{1}  & x_{2}   & x_{3}   & \cdots  & x_{k}   \\
            x_{2}  & x_{3}   & x_{4}   & \cdots  & x_{k+1} \\
            \vdots & \vdots  & \vdots  & \ddots  & \vdots  \\
            x_{l}  & x_{l+1} & x_{l+2} & \cdots  & x_{m}
 			\end{bmatrix}$};
			
\end{tikzpicture}}%
	\caption[Illustration of the Hankel matrix.]{The trajectory matrix $H$ is composed by a series of lagged vectors of size $l$. Due to its structure, this matrix is also known as Hankel matrix.}
	\label{fig:trajectory}
\end{figure}
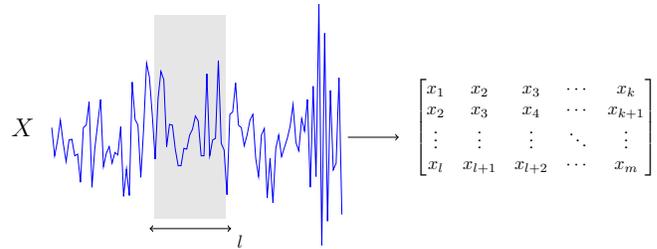
In the literature, the methods can be roughly divided into three categories: (i) deep neural networks, (ii) handcrafted feature extraction followed by a classifier such as support vector machine (SVM) or $k$-Nearest Neighbor ($k$-NN) and (iii) pre-trained feature extraction followed by a classifier. These three categories present benefits and limitations according to the dataset configurations and the application context.

Usually, deep learning methods require a substantial amount of labeled training data. Ko {\it et al.}~\cite{pre_trained} combined multiple pre-trained convolutional neural networks (CNNs) to circumvent this issue. The method works concatenating features produced by multiple pre-trained CNNs followed by dimensionality reduction using linear discriminant analysis (LDA). Finally, an SVM is used for classification. The method can classify sounds of anuran, bird, and insect species, outperforming two types of CNN architectures in terms of overall accuracy. In this method, the memory required depends on the number of pre-trained networks, which may prevent its utilization on low-cost hardware.

A framework based on matrix factorization for bird activity detection and species classification was proposed by Thakur and Rajan~\cite{deep_arch}. The framework joins the properties of matrix factorization with the discriminative capabilities of kernel methods, providing a robust method for bioacoustic signal classification. The archetypal analysis is employed for matrix decomposition, which factorizes an input matrix into a dictionary of archetypes and convex-sparse representations, modeling data boundaries. Also, a deep learning variant of the framework is developed (deep archetypal analysis). Three layers provide improvement in classification accuracy in experimental results using various bioacoustic datasets. Although its competitive accuracy, this approach cannot handle signals of arbitrary size, and all pre-processing steps increase the demand for computing resources.


A solution for insect species classification using the sounds of their wingbeats was proposed by Ntalampiras~\cite{hmm}. The authors present a solution based on a directed acyclic graph (DAG) scheme, wherein the nodes are equipped with a Hidden Markov Model (HMM) for classification. It is claimed that this strategy reduces the problem space. Consequently, it does not require a large amount of data for training, which provides a competitive solution for small bioacoustic datasets. One of the main advantages of this method is that it does not require retraining the model when sounds of new insect species are available. Besides, the method provides interpretability, since the sequence of edges activated in the DAG can be quickly inspected. Since this method is based on an instance-based learning strategy, the model may suffer from overfitting, considering that no regularization scheme is defined.

Nolasco {\it et al.}~\cite{bee_review} provided a solution for classifying beehive states using machine learning and audio data. The data used in this study were obtained as part of the NU-Hive project, aiming to develop a system to monitor beehives' conditions by exploiting the sounds that bees emit. Since bees are the most important pollinators of food crops on the planet, their survival is of high interest. Recently, bee colonies have been declining, an issue that could have drastic consequences for the sustenance of humans and other animals in the food chain. In their study, the authors compare SVM and CNNs to identify the states of the different beehives. One of the most important findings of this work is that SVM was found to generalize better on unseen beehives than CNN when employing features based on Hilbert-Huang Transform (HHT) and Mel-Frequency Cepstral Coefficients (MFCC). Despite its good results, this approach cannot handle signals of arbitrary size, and all pre-processing steps increase the demand for computing resources.

Mutual Singular Spectrum Analysis (MSSA)~\cite{mssa}, also known as Singular Spectrum Classifier, is a classification framework that operates by using subspaces to represent bioacoustic signals, which are generated by employing basis vectors extracted with SVD from trajectory matrices (Fig.~\ref{fig:trajectory}). This approach demonstrated to be efficient in representing and classifying anuran species from their calls. The advantages of this method include its highly compact representation and fast processing time. Since the trajectory matrix can be computed from signals of any length, the proposed framework can handle signals of different sizes without length normalization. Besides, MSSA requires no pre-processing (e.g., segmentation, noise reduction, or syllable extraction), enabling its application on real-time applications under limited hardware conditions.

Although MSSA provided a new signal representation based on subspaces, which is compact and requires no cost-intensive pre-processing techniques, it does not have a discriminant mechanism to extract features aiming at a classification task, since the bioacoustic subspaces of different classes are extracted independently. This drawback impairs the capture of intra-class compactness as well as inter-class separability. To address this issue, Grassmann Singular Spectrum Analysis (GSSA) was proposed~\cite{gssa}. GSSA preserves the advantages of MSSA and improves the robustness of the method by mapping the subspaces onto a Grassmann manifold. The validity of GSSA was shown on the anuran dataset. Some shortcomings of traditional kernel learning~algorithms are presented in the Grassmann manifold. For instance, the computational cost of constructing the kernel grows exponentially with the number of samples to satisfy Mercer's theorem and validate the reproducing kernel Hilbert space.


Knight {\it et al.}~\cite{pre_trained2} employed an AlexNet CNN to classify spectrograms of bioacoustic signals of a bird dataset, and the mean classification accuracy achieved by AlexNet ranged from $88\%$ to $96\%$, according to the parameter configuration used to produce the spectrograms. The best classification accuracy was found when a compound of four spectrograms with distinct scales for frequency, amplitude, and fast Fourier transform (FFT) window size was employed. According to the results, bioacoustic signal classification benefits from selecting the parameters used to convert each audio sample to a spectrogram. One limitation of this method is that it requires a fixed input size, leading to information loss.

A solution for the classification of migrating birds' flight calls based on the fusion of shallow and deep learning features was proposed by Salamon and Bello~\cite{fusion}. The authors investigated an unsupervised dictionary learning based on the spherical $k$-Means algorithm in addition to a CNN. A data augmentation strategy was adopted to deal with the scarcity of training data. The results have shown that the proposed models outperformed MFCC baselines. A late fusion strategy was also used to aggregate shallow and deep features, which improved the classification accuracy by about $2\%$. Despite its good results, this approach cannot handle signals of arbitrary length, and all pre-processing steps increase the demand for computing resources, increasing the hardware cost.

The methods reviewed in this section make extensive use of subspace-related concepts (e.g., LDA and SSA), handcrafted feature extraction, and CNN. Since our objective is to examine the applicability of discriminative subspaces to represent bioacoustic signals, this review provides a general overview of different methods in the literature. By exploiting ideas developed recently in the subspaces theory, the proposed method is described in the next section.
\section{Proposed Method}
\label{sec:proposed}

Throughout the paper, we use the following notation and conventions. Scalars are denoted by lowercase letters and matrices are denoted by uppercase letters. Calligraphic letters are assigned to subspaces and Greek letters are assigned to eigenvectors and canonical angles. The subspace $\mathcal{S}$ spanned by the set of basis vectors $\{\phi_j \in \mathbb{R}^{l} \}_{j=1}^d$ is $d$-dimensional. Given a Hankel matrix $H \in \mathbb{R}^{l \times k}$, $H^{\top}$ denotes its transpose.

Let us consider a classification problem with a dataset containing supervised signals $\{X_i, y_i\}_{i=1}^n$ where $y$ belongs to one of the $c$ classes. In DSSC, the supervised signals are represented by subspaces, and a discriminative space $\mathcal{D}$ is computed based on the estimated class subspaces. The discriminative space $\mathcal{D}$ provides essential information for classification.
Now, given an input signal $X$, its subspace $\mathcal{P}$ is projected onto $\mathcal{D}$ to extract informative features. The projected subspace is then evaluated regarding its distance to the reference subspaces using the canonical angles. The canonical angles will provide the prediction of a label to $X$.

\begin{figure}[htpb!]
	\centering
	\scalebox{.55}{%
	\begin{tikzpicture}
\begin{axis}[
    xlabel={Eigenvectors [$\phi$]},
    ylabel={Eigenvalues [$\lambda$]},
    xmin = -2, xmax = 102, 
    ymin = -0.2, ymax = 2.2,  
    xtick={1, 5, 10, 15, 20, 25, 30, 35, 40, 45, 50, 55, 60, 65, 70, 75, 80, 85, 90, 95, 100},
    ytick={0, 0.2, 0.4, 0.6, 0.8, 1.0, 1.2, 1.4, 1.6, 1.8, 2.0},
    legend pos=north east,
    ymajorgrids=true,
    xmajorgrids=true,
    grid style=dashed,
    height = 8cm,
    width = 14cm,
    axis line style={-|}]
]

\addplot[
    ycomb,
    color=blue,
    mark=*,
    ]
    coordinates {
    (1,1.99996613)(5,1.999898395)(10,1.999695215)(15,1.999085923)(20,1.997260274)(25,1.991803279)(30,1.975609756)(35,1.928571429)(40,1.8)(45,1.5)(50,1)(55,0.5)(60,0.2)(65,0.071428571)(70,0.024390244)(75,0.008196721)(80,0.002739726)(85,0.000914077)(90,0.000304785)(95,0.000101605)(100,3.38696E-05)
    };

    \addplot[
    color=red,
    ]
    coordinates {
    (1,1)(100,1)
    };
    
    \addplot[
    color=black,
    line width=0.1pt,
    dashed,
    ]
    coordinates {
    (50, 2.2)(50, -0.2)
    };

\end{axis}
\end{tikzpicture}

}%
	\caption[Distribution of the eigenvalues of a typical sum subspace $\mathcal{S}$.]{Distribution of the eigenvalues of a typical sum subspace $\mathcal{S}_{(2)}$ for $d = 50$ and $\operatorname{dim}({G}_{(2)}) = 100$.}
	\label{fig:eigen_value}
\end{figure}
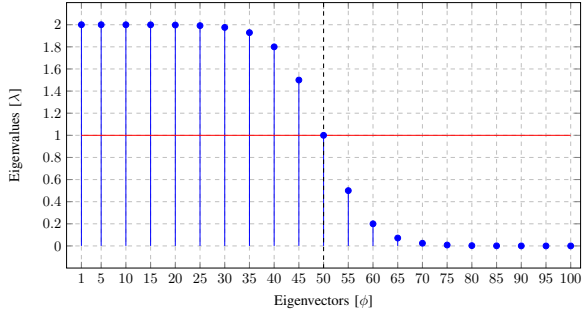

In subspace analysis~\cite{mssa,gssa}, the subspace representation is obtained by the singular value decomposition (SVD) of the trajectory matrix. This new representation provides a high compactness ratio and allows the comparison of nonuniform signal lengths, which is one of the main drawbacks of traditional methods~\cite{len_1,len_2}. Despite such benefits, subspace representation may not be optimal for the classification of bioacoustic signals. The subspaces are obtained independently, neglecting the correlation that may exist between signals belonging to different classes. For illustration, two signals collected from the vocalization of two different species may have their most discriminative features located on minor components of their subspaces, which are usually discarded when basis vectors are selected, permanently impairing this representation. Therefore, we should incorporate a discriminative transformation that preserves the relation between the bioacoustic signals. By applying this transformation, we expect that the distance between subspaces of distinct classes increases, just as decreasing the distance of similar classes, improving the matching of bioacoustic~signals. Fig.~\ref{fig:eigen_value} shows the distribution of the eigenvalues of a sum subspace (e.g.,  $\mathcal{P}_1+\mathcal{P}_2$). The discriminative information is accumulated on the eigenvectors associated with the smallest eigenvalues.

\subsection{SSA for bioacoustic subspace representation}

SSA works by decomposing a signal into independent components. These components can represent trends, periodic components, or noise, depending on the process that generated the signal. SSA consists of two stages, decomposition, and reconstruction. The first stage divides the signal and the second stage rebuilds the decomposed series to provide an enhanced signal. In this work, we are interested in the decomposition properties presented by SSA.

\subsubsection{Creating the trajectory matrix}

First, SSA transforms an input signal $X\in\mathbb{R}^{m}$ into a matrix structure. This procedure is conducted by selecting a vector of $l$ consecutive sub-signals from $X$ and moving this selection throughout the input signal, as shown in Fig.~\ref{fig:trajectory}. This operation can also be regarded as a time embedding and results in the trajectory matrix $H$ with dimensions $l$ by $k$, where these are the maximum autocorrelation time-lag and the length of the time window respectively. The length of the window is determined by the relation $k = m - l + 1$. The procedure of embedding $X$ into its time-delayed coordinates results in a sequence of lagged vectors. This set of lagged vectors is arranged as columns of a trajectory matrix with a Hankel structure, as follows:
\begin{equation}
\small{
H = \left[
\begin{array}{ccccc}
x_{1}  & x_{2}   & x_{3}   & \cdots  & x_{k}   \\
x_{2}  & x_{3}   & x_{4}   & \cdots  & x_{k+1} \\
\vdots & \vdots  & \vdots  & \ddots  & \vdots  \\
x_{l}  & x_{l+1} & x_{l+2} & \cdots  & x_{m}      
\end{array}
\right]~.
\label{eqtraj}
}
\end{equation}
During the decomposition stage, a maximum time lag $l$ should be set, which is usually experimentally obtained (unless strong assumptions are made), and it depends on the signal structure. A useful strategy is to set $l$ proportional to the signal's periodicity to get well-separated components but never higher than $m/2$. Usually, $l \ll k$. If more specific information about the signal is available, the Nyquist rate may provide clues about how to set $l$ appropriately. In this sense, a rule of thumb is to choose $l$ between $\nicefrac{f_s}{20}\leq l\leq\nicefrac{f_s}{10}$, where $f_s$ is the sampling frequency of the bioacoustic records~\cite{COLONNA201864}.

By computing the correlations between the entries of $H$, one can obtain a matrix $U$ whose columns form an orthogonal basis of the $l$-dimensional space. The $l \times l$-dimensional auto-correlation matrix $A$ is obtained as follows:
\begin{equation}\label{eq:correlation}
    A = HH^{\top}~,
\end{equation}
\noindent and the eigenvalue decomposition of $A$ is:
\begin{equation}\label{eq:EVD}
    A = {U}{\Sigma}{U}^{\top}~,
\end{equation}
\noindent where $U$ is the matrix of basis vectors and ${\Sigma}$ $=$ $\operatorname{diag}(\sigma_{1},\cdots,\sigma_{l})$ are the corresponding eigenvalues.
The above decomposition can be used to represent the corresponding bioacoustic signal $X$ with the advantage that this new representation presents the most representative components of the signal in an orderly fashion, facilitating the selection of the most relevant ones for representation.

\subsubsection{Selecting the bioacoustic subspace dimension}

We divide $U$ into two sets: $\overline{U} = \{\phi_k\}_{k=1}^{p}$ and its complement, $\underline{U} = \{\phi_j\}_{j=p+1}^{l}$, to select the most representative basis. The first ${p}$ elements which approximate the original matrix $H$ are employed to span the bioacoustic $p$-dimensional subspace $\mathcal{P}$, compactly representing the bioacoustic signal ${X}$, while the remaining basis vectors are considered as noise. The following ratio measures the contribution of the first ${p}$ elements of $U$ in terms of the reconstruction error of $H$:

\begin{equation}\label{eq:ratio}
    \mu(p) = {\sum_{k=1}^{p}\sigma_k}/{\sum_{j=1}^{l}\sigma_j}~,
\end{equation}

\noindent where $\sigma_j$ is the eigenvalue associated with the $j$-th column of $U$. The subspace $\mathcal{P}$ spanned by the basis vector $\overline{U}$ can compactly represent $X$ regardless of its length. This means that $X$ may have virtually any finite length, which will not change the dimension of $\mathcal{P}$ and $\mu(\cdot)$ controls the trade-off between the reconstruction error of $\mathcal{P}$ and its dimensionality.

It is worth mentioning that the basis vectors $\overline{U}$ and $\underline{U}$ are also known as the principal and minor components. Although these basis vectors are frequently employed for feature extraction and dimensionality reduction, we employ $\overline{U}$ directly for representing an input signal, without projecting $X$ onto $\mathcal{P}$. Both $X$ and its projection are no longer required after obtaining $\overline{U}$, providing memory efficiency.

\subsection{Canonical angles between bioacoustic subspaces}

The canonical angles between two bioacoustic $p$-dimensional subspaces $\mathcal{P}_1$ and $\mathcal{P}_2$ can be calculated by the singular values of ${W}$, which is given by:
\begin{equation}\label{eq:project_sim}
    {W} =  {\overline{U}_{1}}^{\top}{\overline{U}_{2}}~,
\end{equation}
\noindent where the basis vectors $\overline{U}_1$ and $\overline{U}_2$ span the bioacoustic subspaces $\mathcal{P}_1$ and $\mathcal{P}_2$, respectively. Once equipped with the singular values of ${W}$, $\{\delta_k\}_{k=1}^{p}$, the canonical angles can be obtained by:
\begin{align}
  \Delta(\mathcal{P}_1,\mathcal{P}_2) &= \{\theta_1, \theta_2, \ldots, \theta_{p}\} \\
   &=  \{\cos^{-1} (\delta_1), \cos^{-1} (\delta_2), \ldots, \cos^{-1} (\delta_{p})\}
\end{align}

\noindent where the first canonical angle $\theta_1$ is the smallest angle between the subspaces spanned by the basis vectors ${U}_1$ and ${U}_2$. Then, $\theta_2$ is the second smallest angle in the orthogonal direction of $\theta_1$. The canonical angle $\theta_3$ is in an orthogonal direction to both $\theta_1$ and $\theta_2$. The remaining angles follow this rule recursively.

When the elements of $\Delta(\cdot)$ approach zero, the two bioacoustic subspaces are completely overlapped and, therefore, may represent the same signal. On the other hand, when the elements of $\Delta(\cdot)$ approach $\pi/2$, it may be evidence that the signals are uncorrelated.

\subsection{Similarity between two bioacoustic subspaces}

A reasonable method for estimating the similarity between two $p$-dimensional subspaces is by averaging the sum of the canonical angles. This procedure can be achieved as follows:
\begin{equation}\label{eq:sim}
    \gamma({\mathcal{P}_1,~\mathcal{P}_2}) = \frac{1}{p} \sum_{j=1}^{p} \cos^2(\theta_j)~.
\end{equation}
The average of the canonical angles $\gamma({\cdot,\cdot})$ provides interpretability, since $\gamma({\cdot,\cdot})$ approaches $1$ when the bioacoustic subspaces have a large amount of common periodic components, indicating that these subspaces have very high similarity. Therefore, these angles also characterize the similarity between autocorrelation matrices of the signals and, consequently, the similarity between their main frequencies. On the other hand, $\gamma({\cdot,\cdot})$ approaches zero when these subspaces present uncorrelated structures, suggesting that these subspaces represent distinct bioacoustic classes with different main frequencies. One of the advantages of using the canonical angles to define a measure of similarity is their flexibility in expressing the similarity among the oscillatory components contained in $\mathcal{P}_1$ and $\mathcal{P}_2$.

Other applications may further exploit the similarity framework offered by the canonical angles. For instance, in specific applications, it may be beneficial to employ a weighting system where the first canonical angle receives higher importance than the remaining ones. In other applications, the last canonical angle may provide more discriminative information, and therefore higher weight should be assigned to it.

\subsection{Discriminative Singular Spectrum Classifier (DSSC)}
In a multiclass problem, $\{\mathcal{P}_{i}\}_{i=1}^{n}$ is the set of reference bioacoustic subspaces spanned by $\{{U}_{i}\}_{i=1}^{n}$. Then, a subspace $\mathcal{D}_{(n)}$ that can act on $\mathcal{P}_i$ can be developed to extract discriminative information. In DSSC, this procedure can be carried out through the GDS projection, which removes the principal subspace that represents the intersection between the different class subspaces. Thus, we compute a discriminant subspace that preserves only the fundamental components for classification. The normalized sum ${G}_{(n)}$ of the autocorrelation matrices of the $n$ bioacoustic subspaces is computed as follows:
\begin{equation} \label{eq:sum_subspace} 
{G}_{(n)} = \frac{1}{n} \sum_{i=1}^n {U}_{i} {U}_{i}^{\top}~.
\end{equation} 
\noindent Since the matrix ${G}_{(n)}$ has information regarding all the $n$ bioacoustic subspaces, it is interesting to exploit it to extract discriminative elements. We can decompose ${G}_{(n)}$ as follows:
\begin{equation}\label{eq:svd_G}
{G}_{(n)} = {B}\varLambda_{(n)}{B}^{\top}~,
\end{equation}
\noindent where the subset of $B$, denoted as $B^{\star} = \{\psi_k\}_{k=d}^{l}$, which is associated with the smallest eigenvalues $\varLambda_{(n)}$ preserves most of the discriminative information contained in $G_{(n)}$ and can be used to generate the discriminative subspace $\mathcal{D}_{(n)}$. The optimal subspace dimension $d$ is set experimentally by maximizing the degree of orthogonality among the bioacoustic subspaces of all classes projected on $\mathcal{D}_{(n)}$. According to Fukui and Maki~\cite{diff}, the sum subspace $\mathcal{S}_{(n)}$, spanned by ${B}$, is composed of vectors contained in all $\{\mathcal{P}_{i}\}_{i=1}^{n}$, in addition to their linear combinations. Once obtained the sum subspace $\mathcal{S}_{(n)}$, it can be further decomposed in such a way that the principal subspace $\mathcal{F}_{(n)}$ and the difference subspace $\mathcal{D}_{(n)}$ can be put into evidence. The following equation exposes this idea:

\begin{equation}\label{eq:decomposition}
\mathcal{S}_{(n)} = \mathcal{F}_{(n)} \oplus \mathcal{D}_{(n)}~,
\end{equation}

\noindent where $\oplus$ stands for the decomposition of the subspace $\mathcal{S}_{(n)}$ into subspaces $\mathcal{F}_{(n)}$ and $\mathcal{D}_{(n)}$. The above decomposition can be accomplished by analyzing the eigenvalues associated with the eigenvectors spanning the sum subspace. By discarding the eigenvectors associated with the eigenvalues of larger variances, we preserve the discriminative eigenvectors, achieving quasi-orthogonality.

\subsection{Projecting the bioacoustic subspaces onto $\mathcal{D}_{(n)}$}
\label{sec:projecting}
Once equipped with the discriminative subspace $\mathcal{D}_{(n)}$, we can accomplish discriminative structures from $\{\mathcal{P}_{i}\}_{i=1}^{n}$. According to Fukui and Maki~\cite{diff} and Tan et al.~\cite{gods}, this procedure can be performed by carrying two different approaches. The first approach includes projecting subspaces onto a discriminative space, then orthogonalizing the projected subspaces using the Gram-Schmidt orthogonalization. The second procedure involves projecting $X$ onto a discriminative space directly, then applying SVD to generate the projected subspaces. In~\cite{diff} and \cite{gods} are established that these two procedures are algebraically equivalent. In this work, we employ the first procedure since it is computationally more efficient. Therefore, the procedure to compute the basis vectors $\{\dot{{U}}_{i}\}_{i}^{n}$ that span $\{\dot{\mathcal{P}}_{i}\}_{i}^{n}$ is:
\begin{equation} \label{eq:project}
\dot{U}_{i} = \operatorname{orth}\left({B^{\star}}^{\top} U_i\right),
\end{equation}
\noindent where $\operatorname{orth}(\cdot)$ denotes the ortho-normalization of a set of vectors by using the Gram-Schmidt process.

\subsection{Orthogonality degree between biacoustic subspaces}
The Fisher score~\cite{Fisher_score} is broadly employed for model selection and consists of scoring a nested model according to its discriminative importance. More precisely, the Fisher score evaluates the subspace spanned by the selected model regarding the distances between data points of different classes and the distances between data points within the same class. Accordingly, a high Fisher score ensures high inter-class and low intra-class variability, which is desirable. Since this work employs subspaces to represent bioacoustic signals, we present Fisher's formulation in terms of bioacoustic subspaces. Given the discriminative subspace $\mathcal{A}$, the average between-class and within-class variability $f_{b}(\mathcal{A})$ and $f_{w}(\mathcal{A})$ are:

\begin{equation}
  f_{b}(\mathcal{A}) = \frac{1}{n} \sum_{i=1}^{n} \gamma\big(\mathcal{K}_i,~\mathcal{K}\big)~
  \label{eq:fisherb}
  \end{equation}

and 

  \begin{equation}
  f_{w}(\mathcal{A}) = \frac{1}{r} \sum_{i=1}^{n} \sum_{j=1}^{n_i} \gamma\big(\mathcal{P}_{ij},~\mathcal{K}_j\big)~,
    \label{eq:fisherw}
\end{equation}

\noindent where $\mathcal{K}_i$ stands for the Karcher mean of the $i$-th class subspace, $\mathcal{K}$ is the Karcher mean of the $\mathcal{K}_i$ subspaces, $n_i$ is the number of subspaces of the $i$-th class and $r=n\cdot n_i$. Finally, $\gamma(\cdot,\cdot)$ measures the similarity between the bioacoustic subspaces (e.g.~\eqref{eq:sim}). Then, $f(\mathcal{A}) = f_{b}(\mathcal{A})/f_{w}(\mathcal{A})$ reflects the orthogonality degree for bioacoustic subspaces since its value is high when the subspaces of different classes approaches the orthogonal status and the same class subspaces are adjacent.

The above formulation provided to describe $f(\mathcal{A})$ is an unbounded measure. The larger the $f(\mathcal{A})$ value, the smaller the within-class scatter than the between-class scatter. $f(\mathcal{A})$ straightforwardly measures how compact each class is compared to how far it is from the other class. Due to its unbounded formulation, we apply a sigmoid function to establish $f(\mathcal{A})$ bounds in the range $(0,~1)$. Therefore, the bounded Fisher score for bioacoustics subspaces is:

\begin{equation}
  f_{s}(\mathcal{A}) = \frac{1}{1 + \mathit{e}^{-f(\mathcal{A})}}~.
    \label{eq:fisher}
\end{equation}

We adopted the sigmoid function due to its monotonic and bounded nature. Although out of the scope of this paper, the sigmoid activation can be interpreted as probabilities. The introduced score will be employed as an evaluation metric to select the optimal dimension of $\mathcal{D}_{(n)}$, which will be associated with the highest orthogonality degree.
In~\eqref{eq:fisher}, we maximize $f_b(A)$ while minimizing $f_w(A)$ leading to maximizing $f(A)$. In DSSC, its optimization process requires only the proper selection of the dimension $d$ of $\mathcal{D}$. We can achieve quasi-orthogonality between the bioacoustic subspaces by generating the appropriate $\mathcal{D}$. Formally, we can obtain $\mathcal{D}$ as follows:

\begin{equation} \label{eq:optimization}
\mathcal{D}^\star  = \argmax f_s(\mathcal{D})~.
\end{equation}
\section{Experimental Results}
\label{sec:experimental}

In the first experiment, we evaluate the parameters of DSSC, such as the window length $l$ and the bioacoustic subspace dimension $p$ that result in the best representation. In the second experiment, we visualize the relationship between the subspaces by using t-SNE. This visualization gives insight regarding DSSC separability, as well as its representation capabilities. Then, we compare the proposed method with existing task-oriented methods. In the last experiment, we visualize the basis vectors produced by MSSA and DSSC to investigate the oscillatory components' behavior.

\subsection{Datasets}

The anuran dataset~\cite{Colonna2016} consists of $60$ recordings of $10$ different species of frogs with varying record lengths collected under noise conditions. The number of records per species ranges from $3$ to $11$. This dataset provides a genuine challenge since the number of samples is limited due to difficulties in cataloging some species. These recordings were recorded with $44.1$ kHz of sampling rate and $32$ bits.

The mosquito wingbeat dataset~\cite{mosquito} comprises $626$ recordings of $20$ different species of mosquitoes. The records reflected the bioacoustic signatures of free-flying mosquitoes and were acquired using the microphone of mobile phones. These signals were acquired at sampling rates ranging from $8$ kHz to $44.1$ kHz and various file formats, depending on the mobile phone. The signals were converted to a WAV format and resampled to $44.1$ kHz. This dataset is very challenging since mobile phones with different specifications were employed to collect the data, and the length of the recordings varies highly.

The NU-Hive dataset~\cite{bee_review} contains $576$ files of $10$ min duration each, resulting in approximately $96$ hours of recordings. The task is to classify whether the bee queen is present or not inside the beehive. The records came from two beehives and periods when the queen bee was present or absent for each beehive. The data were collected continuously with a sampling rate of $32$ kHz, with sensors located inside the hives.

Table~\ref{tab:datasets} summarizes the audio record lengths of the datasets (number of classes, number of samples per class, total recording time, and sampling rate). Publicly available datasets of bioacoustic signals are limited in size due to the high cost of manual labeling.

\begin{table}[]
\centering
\caption{Summary of the investigated datasets.}
\footnotesize
\label{tab:datasets}
\begin{tabular}{lcccr}
\toprule
\multicolumn{1}{c}{Dataset}        & Samples & Classes & Time Length         & Sampling Rate     \\ \midrule
Anuran~\cite{Colonna2016}               & 60      & 10      & 3 $\sim$ 360 sec    & 44.1 kHz           \\
Mosquito~\cite{mosquito}  & 558     & 20      & 1 $\sim$ 438 sec    & 8 $\sim$ 44.1 kHz  \\
NU-Hive~\cite{bee_review}          & 576     & 10      & 10 min              & 32.0 kHz           \\ \bottomrule

\end{tabular}
\end{table}

\subsection{Evaluating DSSC parameters on NU-Hive dataset}

In this experiment, we employ the NU-Hive dataset to evaluate the window length $l$ of the Hankel matrix, which maximizes the accuracy of MSSA and DSSC and the number of basis vectors $p$ necessary for representing a bioacoustic subspace. This analysis is essential to understand the sensitivity of the proposed method concerning the parameter change. Besides, understanding the parameters' behavior is crucial in developing new bioacoustic systems in similar datasets. The dataset was split into training (50\%) and test (50\%) sets.  

\begin{figure}[htpb!]
	\centering
	\scalebox{.6}{
	\begin{tikzpicture}
\begin{axis}[
    axis background/.style={fill=gray!5},
    xlabel={window length [$l$]},
    ylabel={Accuracy [$\%$]},
    xmin = 10, xmax = 200,
    ymin = 20, ymax = 100,
    legend pos=north east,
    ymajorgrids=true,
    xmajorgrids=true,
    grid style=dashed,
    height = 7cm,
    width = 12cm,
    axis line style={-|}],
    style={thick}
]



\addplot[
    color=blue,
    thick,
    ]
    coordinates {(10,55)(11,56.7333)(12,58.4667)(13,60.2)(14,61.9333)(15,63.6667)(16,65.2667)(17,66.8667)(18,68.4667)(19,70.0667)(20,71.6667)(21,73.4242)(22,75.1818)(23,76.9394)(24,78.697)(25,80.4545)(26,81.5417)(27,82.6289)(28,83.7162)(29,84.8034)(30,85.8906)(31,86.7309)(32,87.5713)(33,88.4116)(34,89.252)(35,90.0923)(36,90.3732)(37,90.6541)(38,90.935)(39,91.216)(40,91.4969)(41,91.3292)(42,91.1616)(43,90.994)(44,90.8263)(45,90.6587)(46,90.5522)(47,90.4458)(48,90.3394)(49,90.2329)(50,90.1265)(51,90.0899)(52,90.0533)(53,90.0167)(54,89.9802)(55,89.9436)(56,89.9594)(57,89.9753)(58,89.9911)(59,90.0069)(60,90.0228)(61,90.0579)(62,90.093)(63,90.1281)(64,90.1632)(65,90.1983)(66,90.3055)(67,90.4128)(68,90.5201)(69,90.6274)(70,90.7346)(71,90.9361)(72,91.1376)(73,91.3391)(74,91.5406)(75,91.7421)(76,91.9965)(77,92.2509)(78,92.5054)(79,92.7598)(80,93.0142)(81,93.2485)(82,93.4827)(83,93.7169)(84,93.9511)(85,94.1854)(86,94.3312)(87,94.477)(88,94.6229)(89,94.7687)(90,94.9145)(91,94.933)(92,94.9514)(93,94.9698)(94,94.9882)(95,95.0067)(96,94.8972)(97,94.7876)(98,94.6781)(99,94.5686)(100,94.4591)(101,94.2558)(102,94.0525)(103,93.8492)(104,93.6459)(105,93.4425)(106,93.2005)(107,92.9584)(108,92.7164)(109,92.4743)(110,92.2323)(111,92.0086)(112,91.7849)(113,91.5612)(114,91.3375)(115,91.1138)(116,90.949)(117,90.7842)(118,90.6195)(119,90.4547)(120,90.2899)(121,90.0956)(122,89.9014)(123,89.7071)(124,89.5129)(125,89.3186)(126,88.977)(127,88.6354)(128,88.2939)(129,87.9523)(130,87.6107)(131,87.0926)(132,86.5745)(133,86.0564)(134,85.5383)(135,85.0202)(136,84.6303)(137,84.2405)(138,83.8506)(139,83.4608)(140,83.0709)(141,82.6606)(142,82.2503)(143,81.84)(144,81.4297)(145,81.0194)(146,80.5909)(147,80.1623)(148,79.7337)(149,79.3051)(150,78.8765)(151,78.4744)(152,78.0722)(153,77.67)(154,77.2679)(155,76.8657)(156,76.6041)(157,76.3425)(158,76.0808)(159,75.8192)(160,75.5576)(161,75.2572)(162,74.9568)(163,74.6563)(164,74.3559)(165,74.0555)(166,73.6536)(167,73.2517)(168,72.8498)(169,72.4479)(170,72.046)(171,71.65)(172,71.2539)(173,70.8579)(174,70.4618)(175,70.0658)(176,69.5501)(177,69.0345)(178,68.5189)(179,68.0032)(180,67.4876)(181,67.1719)(182,66.8562)(183,66.5405)(184,66.2248)(185,65.909)(186,65.4554)(187,65.0018)(188,64.5482)(189,64.0946)(190,63.641)(191,63.3254)(192,63.0098)(193,62.6942)(194,62.3786)(195,62.0629)(196,61.5203)(197,60.9776)(198,60.435)(199,59.8923)(200,58.8923)};
    
\addplot[
    color=red,
    thick,
    ]
    coordinates {(10,45.3)(11,47.0833)(12,48.8667)(13,50.65)(14,52.4333)(15,54.2167)(16,55.8667)(17,57.5167)(18,59.1667)(19,60.8167)(20,62.4667)(21,64.2742)(22,66.0818)(23,67.8894)(24,69.697)(25,71.5045)(26,72.6417)(27,73.7789)(28,74.9162)(29,76.0534)(30,77.1906)(31,78.0809)(32,78.9713)(33,79.8616)(34,80.752)(35,81.6423)(36,81.9732)(37,82.3041)(38,82.635)(39,82.966)(40,82.7969)(41,83.0292)(42,82.8616)(43,82.694)(44,82.5263)(45,82.3587)(46,82.2522)(47,82.1458)(48,82.0394)(49,81.9329)(50,81.8265)(51,81.7899)(52,81.7533)(53,81.7167)(54,81.6802)(55,81.6436)(56,81.6594)(57,81.6753)(58,81.6911)(59,81.7069)(60,81.7228)(61,81.7579)(62,81.793)(63,81.8281)(64,81.8632)(65,81.8483)(66,81.9055)(67,81.9628)(68,82.0201)(69,82.0774)(70,82.1346)(71,82.2861)(72,82.4376)(73,82.5891)(74,82.7406)(75,82.8921)(76,83.0965)(77,83.3009)(78,83.5054)(79,83.5598)(80,83.7142)(81,83.8485)(82,83.9827)(83,84.1169)(84,84.2511)(85,84.3854)(86,84.4312)(87,84.477)(88,84.5229)(89,84.5687)(90,84.6145)(91,84.533)(92,84.4514)(93,84.3698)(94,84.2882)(95,84.2067)(96,83.9972)(97,83.7876)(98,83.5781)(99,83.3686)(100,83.1591)(101,82.8558)(102,82.5525)(103,82.2492)(104,81.9459)(105,81.6425)(106,81.3005)(107,80.9584)(108,80.6164)(109,80.2743)(110,79.9323)(111,79.6086)(112,79.2849)(113,78.9612)(114,78.6375)(115,78.3138)(116,78.049)(117,77.7842)(118,77.5195)(119,77.2547)(120,76.9899)(121,76.6956)(122,76.4014)(123,76.1071)(124,75.8129)(125,75.5186)(126,75.077)(127,74.6354)(128,74.1939)(129,73.7523)(130,73.3107)(131,72.6926)(132,72.0745)(133,71.4564)(134,70.8383)(135,70.2202)(136,69.7303)(137,69.2405)(138,68.7506)(139,68.2608)(140,67.7709)(141,67.2606)(142,66.7503)(143,66.24)(144,65.7297)(145,65.2194)(146,64.6909)(147,64.1623)(148,63.6337)(149,63.1051)(150,62.5765)(151,62.0744)(152,61.5722)(153,61.07)(154,60.5679)(155,60.0657)(156,59.7041)(157,59.3425)(158,58.9808)(159,58.6192)(160,58.2576)(161,57.8572)(162,57.4568)(163,57.0563)(164,56.6559)(165,56.2555)(166,55.7536)(167,55.2517)(168,54.7498)(169,54.2479)(170,53.746)(171,53.25)(172,52.7539)(173,52.2579)(174,51.7618)(175,51.2658)(176,50.6501)(177,50.0345)(178,49.4189)(179,48.8032)(180,48.1876)(181,47.7719)(182,47.3562)(183,46.9405)(184,46.5248)(185,46.109)(186,45.5554)(187,45.0018)(188,44.4482)(189,43.8946)(190,43.341)(191,42.9254)(192,42.5098)(193,42.0942)(194,41.6786)(195,41.2629)(196,40.6203)(197,39.9776)(198,39.335)(199,38.6923)(200,37.6923)};
    
    \legend{DSSC, MSSA}
    
\end{axis}
\end{tikzpicture}}
	\caption{Accuracy on the test set of the NU-Hive dataset when $l$ is modified.}
	\label{fig:param_l}
\end{figure}
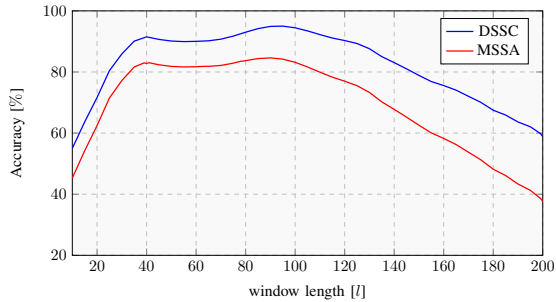

Fig.~\ref{fig:param_l} shows the changes of the accuracy of MSSA and DSSC methods when the window length $l$ varies between $10$ and $200$. The horizontal axis denotes the maximum time lag $l$ used to obtain the Hankel matrix. For this experiment, we set $p$ to account for $90\%$ of the variance of the subspace. From the results, we can verify that the accuracy of MSSA and DSSC increases as $l$ increases until it reaches $40$. After that, a slight drop occurs until $l=60$. The value of $l$ between $90$ and $95$ maximizes both methods' accuracy, leading to $95\%$ and $84\%$ of accuracy for DSSC and MSSA, respectively. This result shows the effect of selecting an appropriate value of $l$ to represent a bioacoustic subspace. When selecting a time lag with a value higher than $100$, the accuracy decreases, suggesting that the main frequencies captured by the autocorrelation were decomposed into non-discriminant signal components, impairing the subspace representation.

\vspace{-20pt}
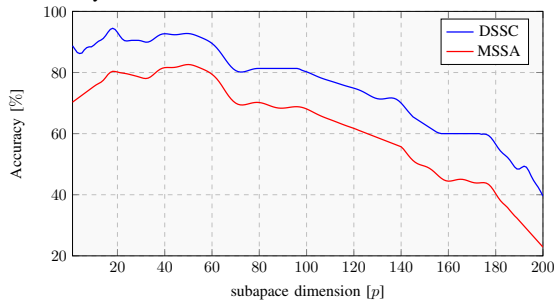
\begin{figure}[htpb!]
	\centering
	\scalebox{.6}{%
	\begin{tikzpicture}
\begin{axis}[
    axis background/.style={fill=gray!5},
    xlabel={subapace dimension [$p$]},
    ylabel={Accuracy [$\%$]},
    xmin = 1, xmax = 200,
    ymin = 20, ymax = 100,
    legend pos=north east,
    ymajorgrids=true,
    xmajorgrids=true,
    grid style=dashed,
    height = 7cm,
    width = 12cm,
    axis line style={-|}],
    style={thick},
    axis background/.style={fill=gray!50}
]

\addplot[
    color=blue,
    thick,
    ]
    coordinates {(1,88.7409)(2,87.6368)(3,86.5519)(4,86.2119)(5,86.3758)(6,87.4295)(7,88.1281)(8,88.3929)(9,88.4002)(10,88.9235)(11,89.6203)(12,90.2343)(13,90.6047)(14,91.1408)(15,92.0586)(16,93.1906)(17,94.1208)(18,94.4657)(19,94.0869)(20,93.1446)(21,91.9925)(22,91.0061)(23,90.4309)(24,90.3151)(25,90.4188)(26,90.5216)(27,90.5237)(28,90.523)(29,90.5134)(30,90.3681)(31,90.124)(32,89.9234)(33,89.9664)(34,90.2655)(35,90.7613)(36,91.3474)(37,91.9063)(38,92.3423)(39,92.6039)(40,92.6906)(41,92.6444)(42,92.5312)(43,92.4193)(44,92.36)(45,92.3765)(46,92.4614)(47,92.5836)(48,92.6993)(49,92.7662)(50,92.7535)(51,92.6488)(52,92.4581)(53,92.2007)(54,91.9013)(55,91.5808)(56,91.249)(57,90.901)(58,90.5173)(59,90.0686)(60,89.5232)(61,88.8551)(62,88.052)(63,87.1198)(64,86.0848)(65,84.9913)(66,83.896)(67,82.8609)(68,81.9447)(69,81.1949)(70,80.6424)(71,80.2976)(72,80.1504)(73,80.1726)(74,80.3224)(75,80.5504)(76,80.8057)(77,81.042)(78,81.222)(79,81.3207)(80,81.33)(81,81.33)(82,81.33)(83,81.33)(84,81.33)(85,81.33)(86,81.33)(87,81.33)(88,81.33)(89,81.33)(90,81.33)(91,81.33)(92,81.33)(93,81.33)(94,81.33)(95,81.33)(96,81.33)(97,81.13)(98,80.7488)(99,80.472)(100,80.2251)(101,79.9235)(102,79.587)(103,79.236)(104,78.8884)(105,78.558)(106,78.2523)(107,77.9727)(108,77.7152)(109,77.4724)(110,77.2356)(111,76.9973)(112,76.7532)(113,76.5028)(114,76.2492)(115,75.9982)(116,75.7555)(117,75.5249)(118,75.3056)(119,75.0914)(120,74.8711)(121,74.6299)(122,74.3531)(123,74.0295)(124,73.6556)(125,73.2385)(126,72.7968)(127,72.3595)(128,71.9615)(129,71.6383)(130,71.418)(131,71.3148)(132,71.3237)(133,71.4179)(134,71.5508)(135,71.6614)(136,71.6837)(137,71.558)(138,71.2416)(139,70.7176)(140,70.0003)(141,69.1337)(142,68.1849)(143,67.2336)(144,66.3569)(145,65.5)(146,65.0)(147,64.5)(148,64.0)(149,63.5)(150,63.0)(151,62.5)(152,62.0)(153,61.5)(154,61.0)(155,60.5)(156,60.2)(157,60.0)(158,60.0)(159,60.0)(160,60.0)(161,60.0)(162,60.0)(163,60.0)(164,60.0)(165,60.0)(166,60.0)(167,60.0)(168,60.0)(169,60.0)(170,60.0)(171,60.0)(172,60.0)(173,60.0)(174,59.7861)(175,59.8571)(176,59.7576)(177,59.3826)(178,58.6803)(179,57.6834)(180,56.5132)(181,55.3451)(182,54.3375)(183,53.553)(184,52.9123)(185,52.2281)(186,51.3269)(187,50.2068)(188,49.1076)(189,48.4264)(190,48.4168)(191,48.9231)(192,49.3397)(193,49.015)(194,47.8058)(195,46.1555)(196,44.7146)(197,43.5365)(198,42.4259)(199,41.0597)(200,39.5375)};

\addplot[
    color=red,
    thick,
    ]
    coordinates {(1,70.2944)(2,70.8635)(3,71.4326)(4,72.0017)(5,72.5708)(6,73.1399)(7,73.709)(8,74.2782)(9,74.8473)(10,75.4164)(11,75.9497)(12,76.3813)(13,76.8047)(14,77.4408)(15,78.3475)(16,79.3572)(17,80.0542)(18,80.3879)(19,80.3313)(20,80.164)(21,79.9583)(22,79.8336)(23,79.7189)(24,79.5804)(25,79.4)(26,79.2)(27,79)(28,78.8)(29,78.5777)(30,78.3365)(31,78.1141)(32,78.0271)(33,78.1664)(34,78.5655)(35,79.1613)(36,79.8474)(37,80.5063)(38,81.0423)(39,81.4039)(40,81.5906)(41,81.6444)(42,81.6312)(43,81.6193)(44,81.66)(45,81.7765)(46,81.9614)(47,82.1836)(48,82.3993)(49,82.5551)(50,82.6091)(51,82.5377)(52,82.3581)(53,82.1007)(54,81.8013)(55,81.4808)(56,81.149)(57,80.801)(58,80.4173)(59,79.9686)(60,79.4232)(61,78.7551)(62,77.952)(63,76.9865)(64,75.8793)(65,74.669)(66,73.4737)(67,72.3609)(68,71.3947)(69,70.5949)(70,69.9924)(71,69.5976)(72,69.4004)(73,69.3726)(74,69.4724)(75,69.6504)(76,69.8557)(77,70.042)(78,70.172)(79,70.2207)(80,70.1769)(81,70.0427)(82,69.8316)(83,69.5662)(84,69.2743)(85,68.9854)(86,68.727)(87,68.5217)(88,68.3846)(89,68.3217)(90,68.3297)(91,68.3964)(92,68.5021)(93,68.6223)(94,68.7302)(95,68.8)(96,68.8101)(97,68.7453)(98,68.5988)(99,68.3722)(100,68.0751)(101,67.7235)(102,67.337)(103,66.936)(104,66.5384)(105,66.158)(106,65.8023)(107,65.4727)(108,65.1652)(109,64.8724)(110,64.5856)(111,64.2973)(112,64.0032)(113,63.7028)(114,63.3992)(115,63.0982)(116,62.8055)(117,62.5249)(118,62.2556)(119,61.9914)(120,61.7211)(121,61.4299)(122,61.1031)(123,60.8)(124,60.5)(125,60.2)(126,59.9)(127,59.6)(128,59.3)(129,59.0)(130,58.7)(131,58.4)(132,58.1)(133,57.8)(134,57.5)(135,57.2)(136,56.9)(137,56.6)(138,56.3)(139,56.0)(140,55.7)(141,54.9337)(142,53.9349)(143,52.9336)(144,52.0069)(145,51.2161)(146,50.5944)(147,50.1415)(148,49.8241)(149,49.5834)(150,49.3484)(151,49.0515)(152,48.6441)(153,48.1072)(154,47.4566)(155,46.7394)(156,46.0239)(157,45.385)(158,44.8872)(159,44.5712)(160,44.4453)(161,44.4836)(162,44.6329)(163,44.8239)(164,44.9865)(165,45.0645)(166,45.026)(167,44.8702)(168,44.6256)(169,44.3437)(170,44.0856)(171,43.9057)(172,43.8344)(173,43.8629)(174,43.9361)(175,43.9571)(176,43.8076)(177,43.3826)(178,42.6303)(179,41.5834)(180,40.3632)(181,39.1451)(182,38.0875)(183,37.253)(184,36.5623)(185,35.8281)(186,34.8769)(187,33.9739)(188,33.0955)(189,32.3505)(190,31.4876)(191,30.6247)(192,29.7618)(193,28.8989)(194,28.036)(195,27.1731)(196,26.3102)(197,25.4473)(198,24.5844)(199,23.7215)(200,22.8586)};
    
    \legend{DSSC, MSSA}
    
\end{axis}
\end{tikzpicture}}%
	\caption{Accuracy on the test set of the NU-Hive dataset as a function of $p$.}
	\label{fig:param_p}
\end{figure}

Fig.~\ref{fig:param_p} shows the effects on the accuracy of the models when the number of basis vectors $p$ ranges between $1$ and $200$. According to the results, DSSC always achieves the best accuracy, and it requires fewer basis vectors than MSSA to achieve the same level of accuracy. This observation confirms that projecting the bioacoustics subspaces in a discriminative space may reveal correlations that were not immediately available, improving the performance of DSSC. In this experiment, we set $l=95$, which was the optimal value found in the previous experiment. The values of $p$ equal to $18$ and $51$ 
maximize the accuracy of both methods producing approximately $96\%$ and $84.5\%$ of accuracy for DSSC and MSSA, respectively. Table~\ref{tab:results_nuHive} summarizes the accuracy of both methods, as well as the optimal values of parameters $l$ and $p$. In practical terms, these two methods demonstrated relative robustness regarding $l$ when compared to changes in the basis vectors $p$. This observation implies that one should tune the number of basis vectors employed to represent the subspaces more carefully than the autocorrelation time lag. The obtained results show the importance of comparing the whole structures of the subspaces by using multiple basis vectors, indicating that this strategy benefits the comparison of the bioacoustic subspaces.

\begin{table}[htpb!]
\centering
\footnotesize
\caption{Results obtained from MSSA and DSSC on the NU-Hive dataset} 

\label{tab:results_nuHive}
\begin{tabular}{@{}llcclcc@{}}
\toprule
\multicolumn{1}{c}{\multirow{2}{*}{Method}} &  & \multicolumn{2}{c}{$p=95\%$ of Variance}       &  & \multicolumn{2}{c}{$l=95$}                     \\ \cmidrule(l){2-7} 
\multicolumn{1}{c}{}                        &  & Optimum $l$ & Accuracy                         &  & Optimum $p$ & Accuracy                         \\ \cmidrule(r){1-7}
DSSC                                        &  & 95          & \textbf{95}$\%$ &  & 18          & \textbf{96}$\%$ \\
MSSA                                        &  & 90          & 84$\%$                           &  & 51          & $85\%$                           \\ \bottomrule
\end{tabular}
\end{table}

\subsection{Separability of MSSA, DSSC and related methods}

In this experiment, we evaluate the discriminative process of DSSC using the mosquito wingbeat dataset. For this aim, we employ the Fisher score as a separability index for bioacoustic subspaces. Fischer score approaches $1.0$ when the distance between the subspaces of different classes is high, and the distance between the same classes subspaces is low. On the other hand, Fischer score approaches $0.0$ when the distance between the subspaces of different classes is low, and the distance between the same classes subspaces is high. We also employ two common descriptors for audio data: Mel Frequency Cepstral Coefficients (MFCC) and Linear Prediction Coefficients (LPC). MFCC is based on the human hearing system with the hypothesis that the human ear is a robust audio recognizer~\cite{MFCC1,MFCC2}. Due to its versatility and precision, MFCC has been widely used in audio applications, including bioacoustic recognition~\cite{Xie2018}. LPC mimics the human vocal tract~\cite{LPC}, producing a reliable audio descriptor. LPC works by estimating the formants, reducing their effects from the speech signal, and determining the residue's intensity and frequency. One of the advantages of LPC is its compact representation, which benefits the encoding of high-quality speech. In contrast to descriptors based on Fourier transform, which assume a superposition of sinusoids as the generative process, LPC assumes that the acoustic system producing the phenomenon is resonant.
 
\begin{figure*}[htpb!]
	\centering
	\subfigure[LPC]{
	\scalebox{.30}{%
	\includegraphics{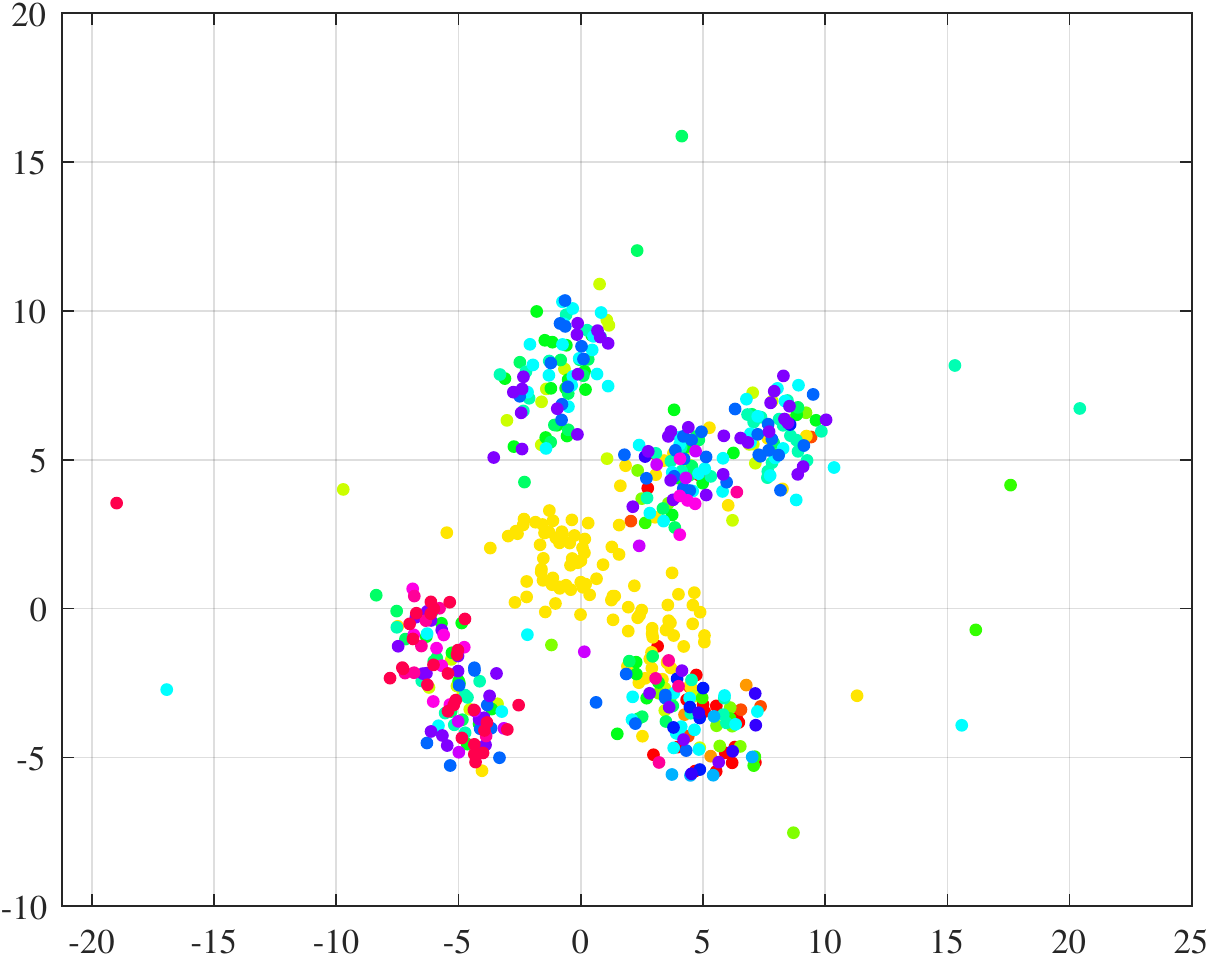}}
	}
	\subfigure[MFCC]{
	\scalebox{.30}{%
	\includegraphics{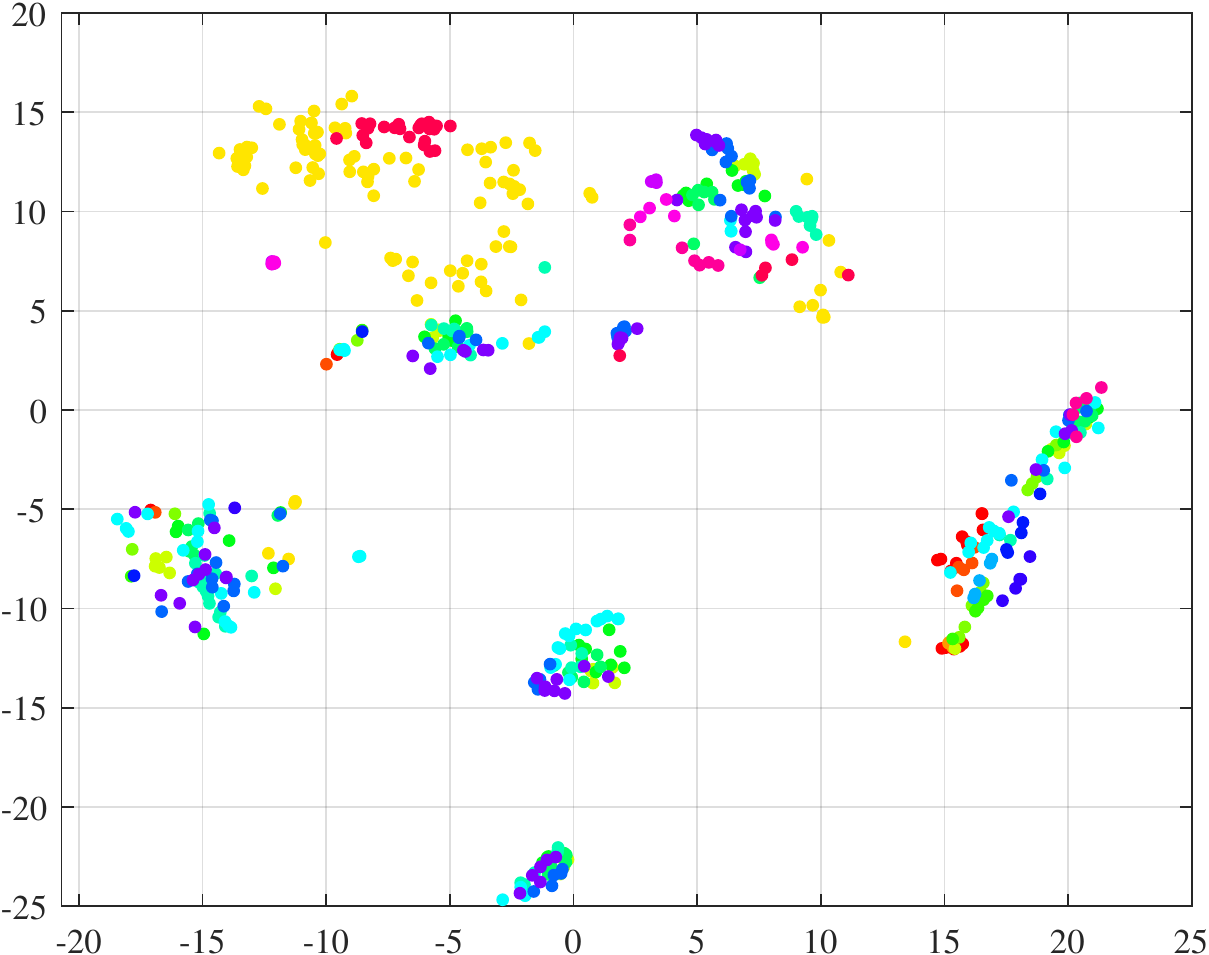}}
	}
	\subfigure[MSSA]{
	\scalebox{.30}{%
	\includegraphics{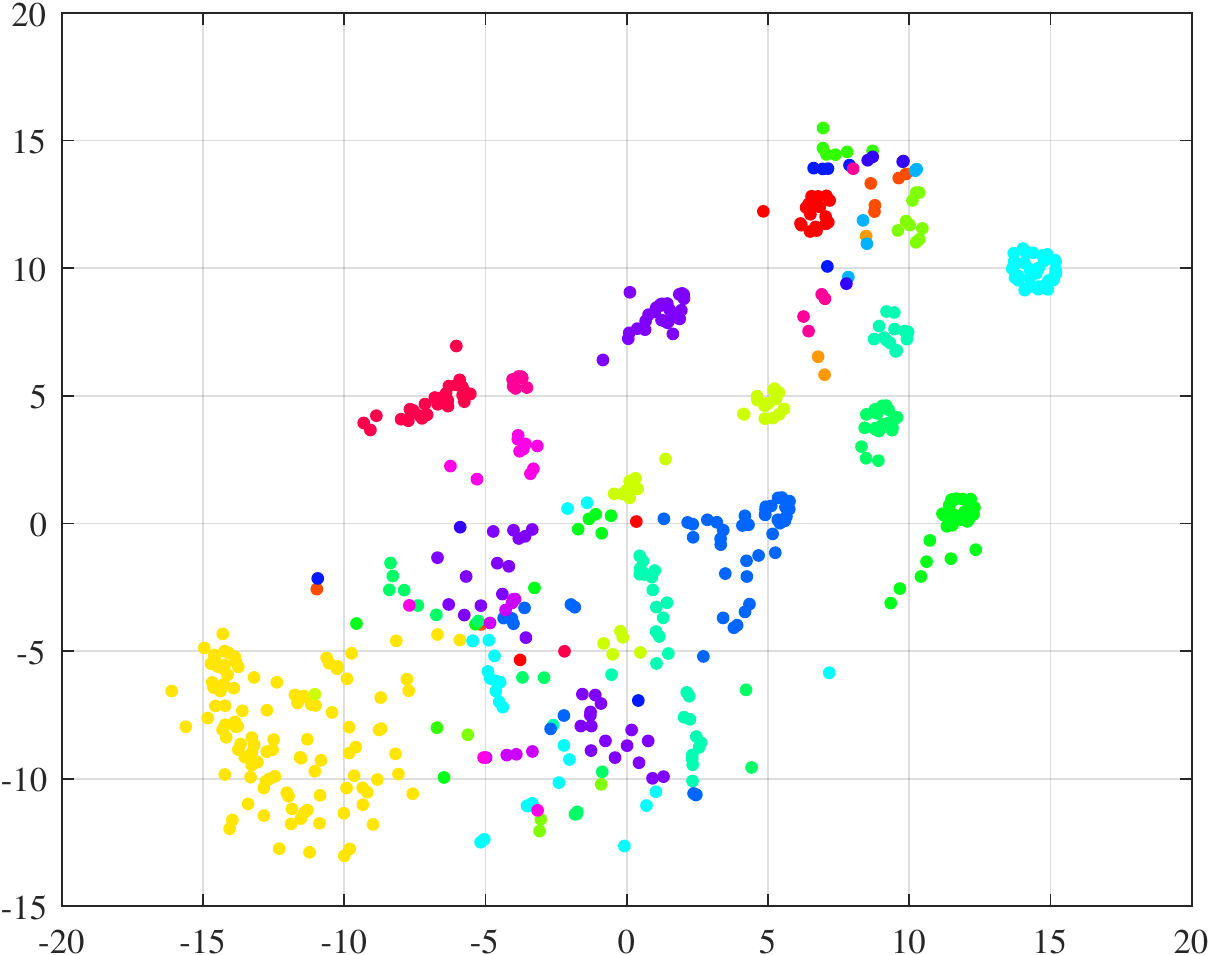}}
	}
	\subfigure[DSSA]{
	\scalebox{.30}{%
	\includegraphics{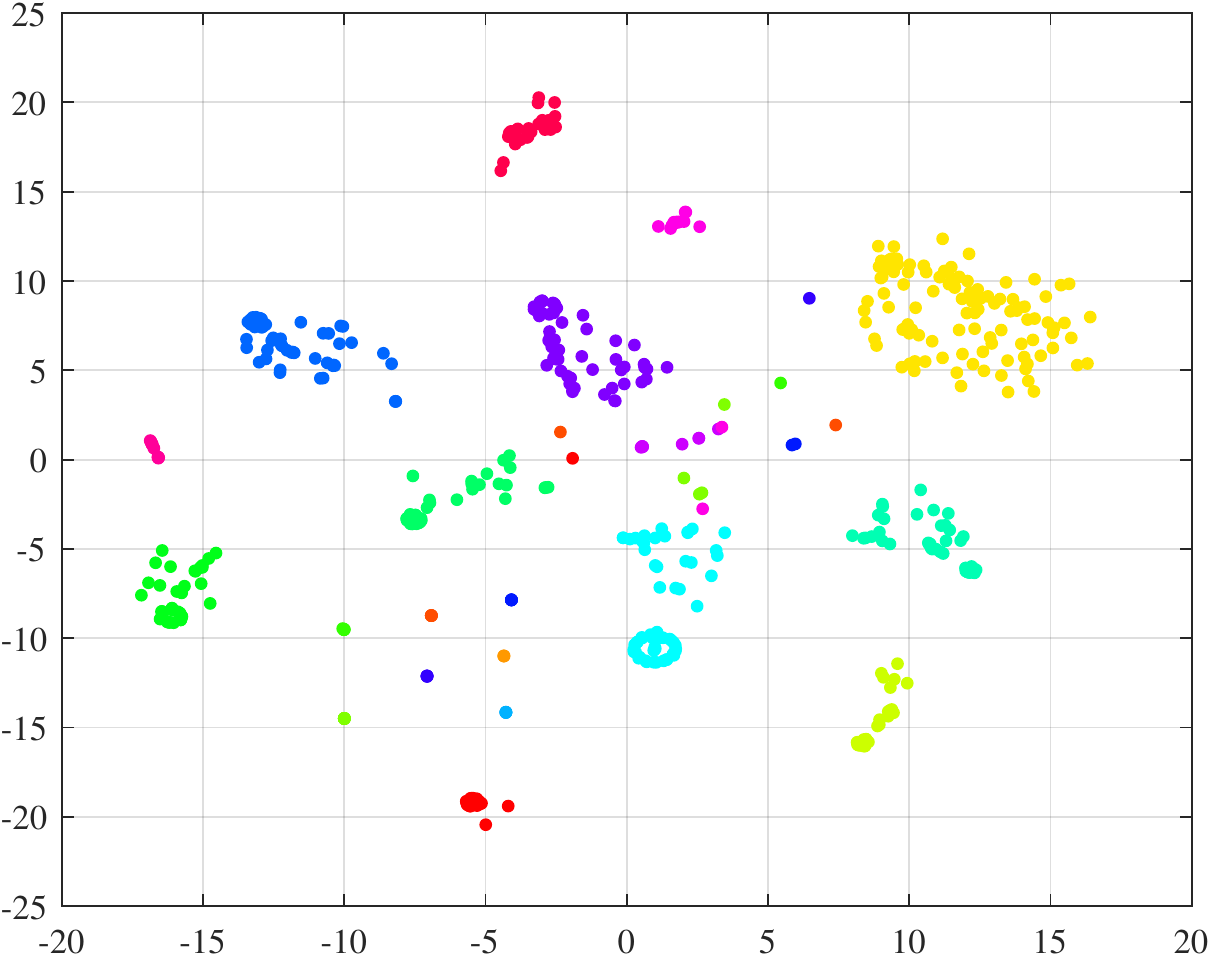}}
}
	\caption{Scatter plots using the t-SNE embedding showing distances between the $20$ classes of the mosquito wingbeat dataset using four different methods. In the plots, the perplexity and epsilon were set to $10$ and $5$, respectively.}
	\label{fig:separability}
\end{figure*}

Here we employ the t-SNE embeddings~\cite{maaten2008visualizing} to visualize the features presented by MSSA, DSSC, MFCC, and LPC. t-SNE is a dimensionality reduction technique that maintains the original high-dimensional data's metric properties and is frequently employed to feature visualization. Fig.~\ref{fig:separability} shows the scatter plots of LPC, MFCC, MSSA, and DSSC. Each point corresponds to one sample from the mosquito wingbeat dataset in the plots, and the different colors denote the different classes. We employed $20$ MFCCs and $12$ LPCs to represent the mosquito wingbeat audio samples since these parameters are commonly used in literature~\cite{Xie2020,Colonna2016}. According to the t-SNE plots, LPC clusters are visually more compact but exhibiting many outliers; differently, the MFCC clusters appear more separable than LPC. Both clusters show a high overlapping among different classes, which may negatively interfere with the classification accuracy.

For the subspace-based methods, we set the window length of $l=200$ and the number of subspaces $p$ to $9$. The dimension $d$ of the discriminative subspace that maximizes the Fisher score for bioacoustic subspaces is $41$. The separability index computed by the Fisher score for bioacoustic subspaces is $0.39$ for MSSA and $0.76$ for DSSC. These indexes indicate that the discriminative mechanism employed by DSSC for bioacoustic subspaces offers more reliable features for classification than the ones provided by MSSA. It is worth mentioning that the Fisher score enforces the class separability, which decreases the overlapping between the different class features. 
According to the results shown by t-SNE, the dispersion in DSSC and MSSA clusters seems to be greater than in LPC and MFCC, and some clusters present elongated shapes. The clusters produced by MSSA and DSSC are visually far more separated than the ones produced by MFCC and LPC. These results show that the bioacoustic signals benefit from the feature extraction representation provided by the subspaces. Besides, the clusters presented by DSSC exhibited a higher separability among the different classes than those produced by MSSA. This suggests that the discriminative mechanism adopted by DSSC provides more reliable clusters than the ones presented by MSSA.

\subsection{Comparison with Task-Oriented Bioacoustic Systems}

The anuran dataset~\cite{Colonna2016} also presents the subject labels.
With this information, it is possible to perform a leave-one-subject-out (LOSO) cross-validation (CV) to evaluate the model's generalization. In this setting, subjects that are not present in the dataset are considered new specimens for testing, which is the case in a realistic application scenario. The LOSO protocol is more challenging for a bioacoustic recognition system and less influenced by background environmental noise, presenting less biased accuracy.

For the comparison purpose, we adopted the task-oriented model developed by Colonna {\it et al.}~\cite{Colonna2016}, which comprises four steps: noise filtering, syllable segmentation, feature extraction, and classification, which is carried out by a standard classification method, such as $k$-NN or SVM. Segmentation and syllable extraction are carried out with the method proposed by Colonna {\it et al.}~\cite{COLONNA20157367}, which is based on the energy of the signal, followed by the extraction of MFCC features. On the other hand, DSSC and MSSA models directly perform signal classification of long-term recordings with several syllables in a single step, resembling an end-to-end system. Nevertheless, as the two systems are assessed using LOSO CV, we can assume that the results are comparable. Thus, we attempt to observe whether the results achieved by the bioacoustic subspace-based methods are competitive in terms of precision, recall, and F-score compared to a more sophisticated and computationally expensive approach. Table~\ref{tab:comparison} shows the performance achieved by $k$-NN and SVM classifiers considering hundreds of syllables since task-oriented systems handle syllables.
These two classifiers were evaluated using the one-against-one binary decomposition strategy, which simplifies multiclass problems and increases accuracy~\cite{roundrobin2001}.
\begin{table}[]
\centering
\footnotesize
\caption{Results obtained from task-oriented systems and E2E-like methods on anuran dataset using LOSO CV.}
\label{tab:comparison}
\begin{tabular}{@{}lcccc@{}}
\toprule
\multicolumn{1}{c}{\multirow{4}{*}{Species}} & \multicolumn{2}{c}{Task-Oriented} & \multicolumn{2}{c}{E2E-like}           \\ \cmidrule(l){2-3}\cmidrule(l){4-5}  
\multicolumn{1}{c}{}                         & $k$-NN            & SVM             & MSSA     & DSSC  \\ \cmidrule(l){2-3}\cmidrule(l){4-5}
\multicolumn{1}{c}{}            & $k$=1           & $p$=3           & $l$=45, $p$=9  & $l$=95, $p$=18  \\
\midrule
\texttt{(a)} \textit{Adenomera andreae}               & 0.33            & 0.30            & 0.60    & \textbf{0.80}    \\
\texttt{(b)} \textit{Ameerega trivittata}             & 0.89            & 0.63            & \textbf{1.00}    & \textbf{1.00}    \\
\texttt{(c)} \textit{Adenomera hylaedactyla}          & 0.98            & \textbf{0.99}           & 0.78    & 0.84    \\
\texttt{(d)} \textit{Hyla minuta}                     & 0.61            & 0.68            & 0.83    & \textbf{0.87}    \\
\texttt{(e)} \textit{Hypsiboas cinerascens}           & \textbf{0.96}            & 0.94            & 0.44    & 0.57    \\
\texttt{(f)} \textit{Hypsiboas cordobae}              & \textbf{1.00}            & \textbf{1.00}            & 0.66    & 0.57    \\
\texttt{(g)} \textit{Leptodactylus fuscus}            & 0.63            & 0.62            & 0.66    & \textbf{1.00}    \\
\texttt{(h)} \textit{Osteocephalus oophagus}          & 0.42            & 0.36            & {\color{black}0.00}    & \textbf{1.00}    \\
\texttt{(i)} \textit{Rhinella granulosa}              & 0.39            & 0.46            & 0.57    & \textbf{0.66}    \\
\texttt{(j)} \textit{Scinax ruber}                    & {\color{black}0.00}            & 0.32            & \textbf{1.00}    & \textbf{1.00}    \\ \midrule 
\multicolumn{1}{l}{Average Precision} & 0.62            & 0.70         & 0.65    & \textbf{0.83}   \\
\multicolumn{1}{l}{Average Recall}    & 0.63            & 0.63         & 0.63    &  \textbf{0.74}  \\
\multicolumn{1}{l}{Average F-score}   & 0.62            & 0.66         & 0.64    &  \textbf{0.78}  \\
\multicolumn{1}{l}{Micro-accuracy} & \textbf{0.86}   & 0.84         & 0.66    &  0.78  \\\bottomrule
\end{tabular}
\end{table}
We highlight in bold the results of the last column corresponding to the species in which DSSC achieved the best results in terms of precision. Given the precision obtained for each anuran species, we can conclude that both MSSA and DSSC are competitive compared to task-oriented systems. The linear combination of the oscillatory components, intrinsic to their subspace formulation, provides robustness to handle datasets with few examples. Additionally, both MSSA and DSSC  inherit the advantages of SSA, such as noise filtering and segmentation-free, in a unified fashion, demonstrating comparable capabilities with task-oriented solutions. However, MSSA and $k$-NN failed to recognize one species, which is unacceptable for a bioacoustic monitoring system.
It is worth noticing that DSSC achieved the worst results for {\it Hypsiboas}, while the $k$-NN achieved the best results. An explanation for this result could be that the difference space $\mathcal{D}_{(n)}$ is not able to recover representative vectors from representing the signal of {\it Hypsiboas}. This behavior is usually observed when a large section of the linear subspace that represents {\it Hypsiboas} is contained in the principal subspace $\mathcal{F}_{(n)}$. Since $\mathcal{F}_{(n)}$ is removed, the projection of this particular subspace onto $\mathcal{D}_{(n)}$ may decrease the representational power of these subspaces instead of improving it.

In general, the proposed DSSC produced the best results among the methods compared. The difference subspace used in DSSC reveals discriminative structures hidden in the oscillatory components existing in the basis vectors of the subspaces, improving the results compared to MSSA. DSSC presented an advantage in average precision, recall, and F-score. Although DSSC did not achieve the best micro-accuracy compared to $k$-NN, the values were not so far. The literature provides evidence that micro-accuracy is not a reliable metric when classes are unbalanced~\cite{metric_1,metric_2}. More precisely, employing standard metrics in imbalanced tasks may provide misleading evidence since these metrics cannot describe skewed domains. In our experiments, $k$-NN handled hundreds of syllables to produce its confusion matrix. Thus, increasing the final value of micro-accuracy.

\subsection{Information captured by the oscillatory components}
\label{exp:oscillatory}

\begin{figure}[]
\centering
\includegraphics[width=0.9\linewidth]{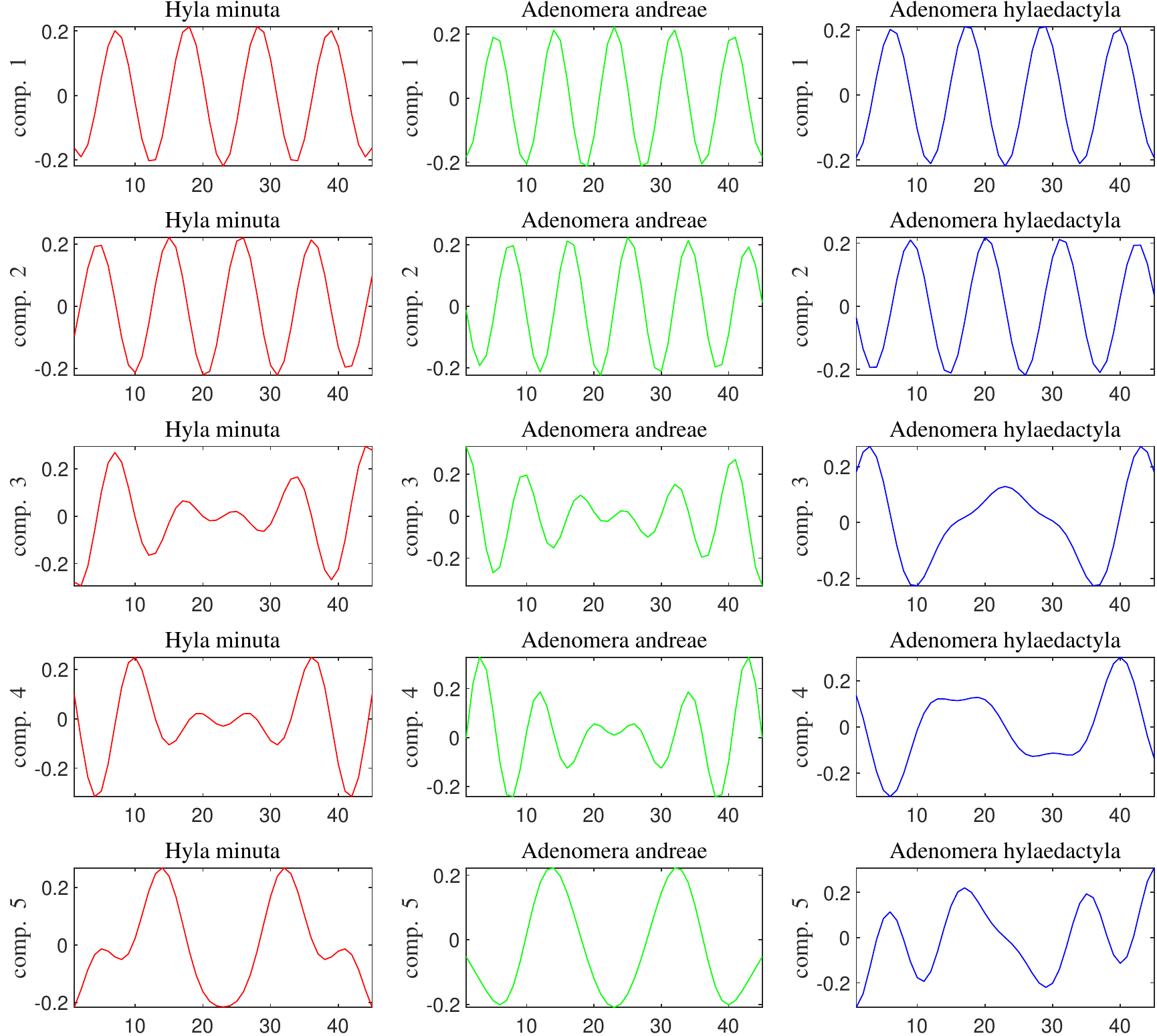}
\caption{First five oscillatory components produced by the MSSA.}
\label{fig:comp_mssa}
\end{figure}

We select the species \textit{Hyla minuta}, \textit{Adenomera andreae} and \textit{Adenomera hylaedactyla} to analyse their oscillatory components since the species \textit{Hyla minuta} was confused with the species \textit{Adenomera andreae} and \textit{Adenomera hylaedactyla}. The supplemental material 
provides the confusion matrices of MSSA and DSSC. Fig.~\ref{fig:comp_mssa} compares the first five eigenvectors of these species. We can notice that the first two oscillatory components are visually identical, although they belong to different species. Even the other three oscillatory components of the species \textit{Hyla minuta} and \textit{Adenomera andreae} are very similar. Since the first oscillatory components are the most important for classification when applying the canonical angles, it is clear that, in this scenario, these features may weaken the classification accuracy of the MSSA. Therefore, a more discriminative mechanism is required.

\begin{figure}[]
\centering
\includegraphics[width=0.9\linewidth]{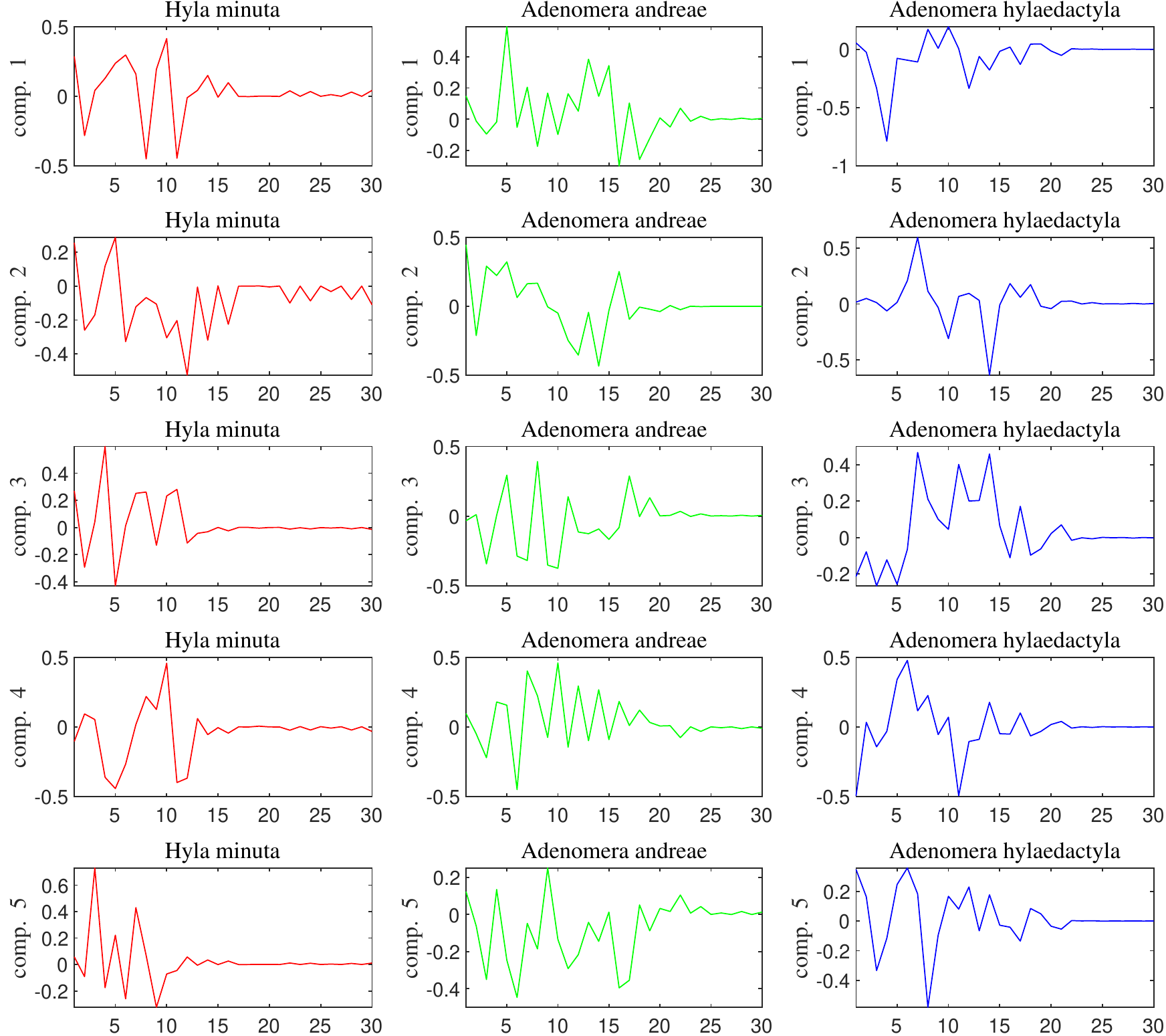}
\caption{First five oscillatory components produced by the DSSC.}
\label{fig:comp_dssc}
\end{figure}

On the other hand, as shown in Fig.~\ref{fig:comp_dssc}, the oscillatory components of the projected subspaces produced by DSSC present discriminant information for classification. The first components of DSSC subspaces are no longer (visually) similar and may benefit the classification accuracy of DSSC. This new representation avoided two misclassifications, which can be seen in supplemental material (row (d) of DSSC's confusion matrix).
This aspect is directly related to the discriminative nature of the difference subspace, which acts by exposing features that are not shared between the bioacoustic classes. More precisely, the discriminative subspace reveals signal structures that improve DSSC classification. According to this observation, we can confirm that bioacoustic subspaces generated by DSSC produce more distinctive features than those provided by MSSA.

\begin{figure}[]
\centering
\includegraphics[width=0.9\linewidth]{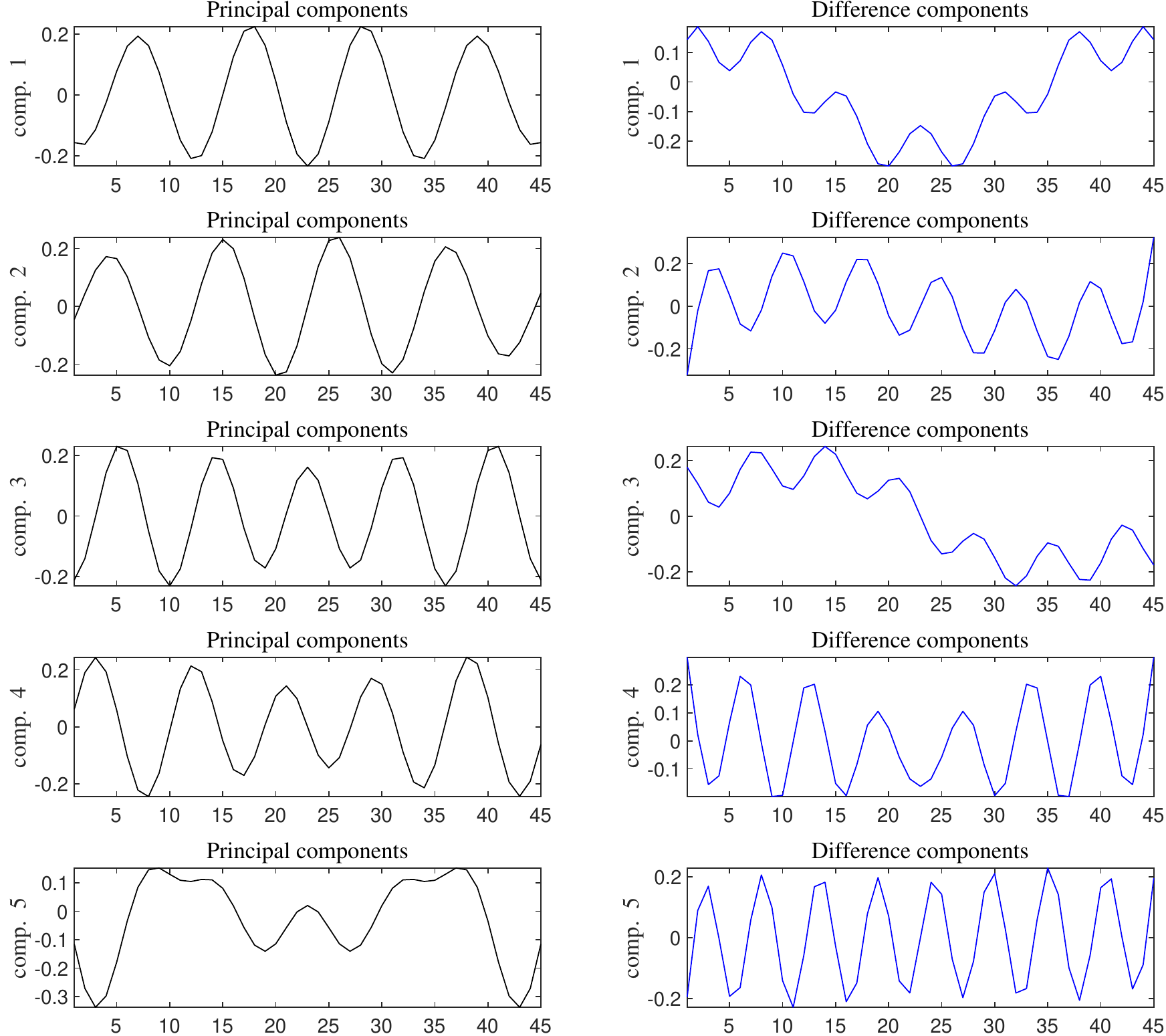}
\caption{First five principal and difference oscillatory components.}
\label{fig:comp_prin_diff}
\end{figure}

In Fig.~\ref{fig:comp_prin_diff}, we can notice that the oscillatory components of the difference subspace exhibit higher variability than the ones provided by the principal space. This indicates that the representation generated by the principal subspace may offer less feature diversity, which may include redundancy. For instance, the first four components of the principal space present visually similar shapes. On the other hand, the oscillatory components of the difference subspace present a richer shape variability, which may extract extra diverse features from the bioacoustic subspace classes.

\section{Conclusions and Future Directions}
\label{sec:conclusion}

We proposed a bioacoustic signal classification method based on signal subspace representation called Discriminative Singular Spectrum Classifier. We developed a discriminative subspace based on the algebraic concept of the difference between subspaces. 
We developed a framework capable of handling bioacoustic signals through this concept, achieving improvements in Beehive, Anuran, and Mosquito datasets. We also analyzed the feature vectors learned from DSSC, confirming that it can extract highly discriminative features from bioacoustic signals without any preprocessing steps.

We evaluated the proposed solution on three different environmental tasks, in which each dataset has a different biological and ecological purpose, yielding the most favorable result in each task. These three tasks have different characteristics from the point of view of signal processing. For instance, the frequency given in the samples of mosquitoes is high and continuous. The frequencies of the signals in the bee samples are lower than the mosquitoes with repetitive patterns. Differently, the anuran presents a richer wave variability, with varying syllable lengths and distinct repetition patterns.

We evaluated the parameters of MSSA and DSSC to understand the classifier behavior. The Beehive dataset was employed for this task, and DSSC outperformed MSSA in this classification task, suggesting that the discriminative mechanism employed by DSSC is an essential tool for bioacoustics feature selection. We then compared the feature separability presented by MSSA and DSSC against commonly used feature extraction techniques. The results presented by t-SNE confirm that DSSC offers an advantage as a feature extraction technique compared to MSSA, LPC, and MFCC.

We evaluated the proposed method with the anuran dataset and compared its results with existing task-oriented methods. The results show that DSSC is superior to MSSA and existing task-oriented methods in most evaluated metrics. In the last experiment, we visualized the basis vectors produced by MSSA and DSSC to investigate the oscillatory components' behavior in both methods. The results show that DSSC can remove common oscillatory components from bioacoustic classes that do not contribute to the classification.

Despite these challenges, the proposed method achieves excellent results in the given tasks, revealing its ability to represent and classify a wide range of bioacoustic signals. Our method shares most of the characteristics seen in E2E bioacoustic systems, such as reduced processing steps, robustness to white Gaussian noise, no segmentation requirements, and automatic feature extraction. DSSC handles signals of varying lengths and achieves higher precision than task-oriented methods. Overall, the segmentation process employed in the task-oriented is handcrafted, requiring technical knowledge regarding the anuran species (for instance) in addition to laborious experiments to validate their assumptions. Differently, MSSA and DSSC do not require such assumptions or technical expertise. All these capabilities are given in a lightweight framework, benefiting remote sensing-related applications.

Although the proposed method might be of interest to biologists studying animal behavior or counting and supervising wildlife, possible direct impact includes representation and analysis of brain signals, breathing phase, and heart rhythm. Since DSSC is based on the autocorrelation matrix, we can assume that our system's application range is not limited to bioacoustic signals only; our system could offer a solution for other signal processing tasks with regular patterns. 

In future work, we aim to exploit nonlinear patterns, which is one limitation of the proposed method. In such an approach, kernel PCA may be used to extract nonlinear patterns and improve the signals' representation.


\ifCLASSOPTIONcaptionsoff
  \newpage
\fi



%
%
%

\bibliographystyle{IEEEtran}
\bibliography{refs}

\vspace{-30pt}
\begin{IEEEbiography}[{\includegraphics[width=1in,height=1.25in,clip,keepaspectratio]{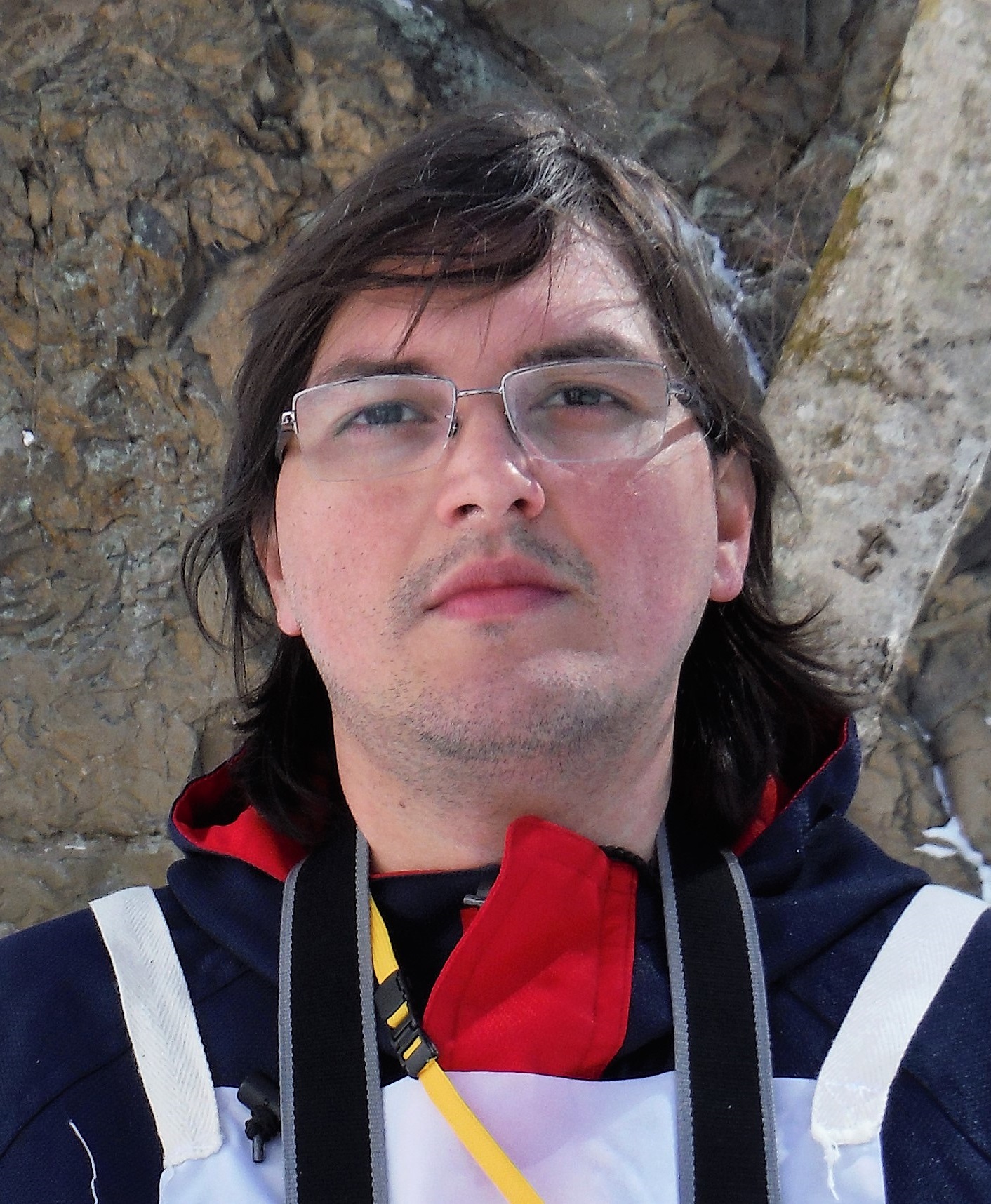}}]{Bernardo B. Gatto} received his B.E. in Computer Engineering in 2010 from Amazonas State University, Brazil, M.E. in Computer Science in 2013 from the University of Tsukuba, Japan and PhD in Computer Science in 2020 from the Federal University of Amazonas. He joined the Center for Artificial Intelligence Research (C-AIR), University of Tsukuba, in 2018. His research interests include digital signal processing, pattern recognition, machine learning and computer vision.
\end{IEEEbiography}
\vspace{-30pt}
\begin{IEEEbiography}[{\includegraphics[width=1in,height=1.25in,clip,keepaspectratio]{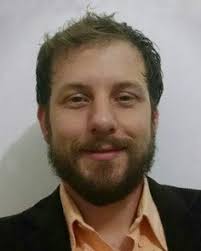}}]{Juan G. Colonna}

is an Associate Professor in the Institute of Computing (IComp) of the Federal University of Amazonas (UFAM). He received the B.Eng. degree in telecommunications engineering from the National University of Río Cuarto, Argentina, in 2009. His M.Sc. and Ph.D. degrees in Informatics were awarded by UFAM in 2013 and 2017, respectively.
He participated as a fellow in a long-term environmental monitoring project with the Mamirauá Institute (2017-2018). Since then, he has worked with computational methods applied to ecological monitoring, mainly related to bioacoustics and ecoacoustics.
\end{IEEEbiography}
\vspace{-30pt}
\begin{IEEEbiography}[{\includegraphics[width=1in,height=1.25in,clip,keepaspectratio]{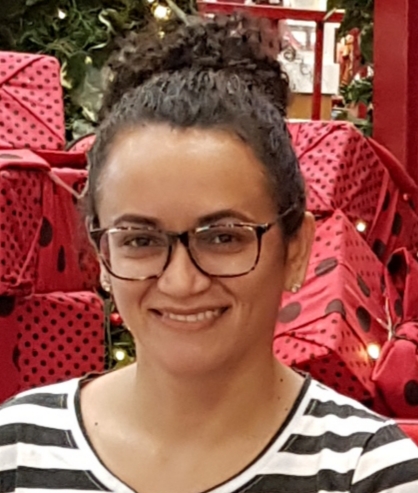}}]{Eulanda M. dos Santos}
is an Associate Professor in the Institute of Computing (IComp) of the Federal University of Amazonas. She received a B.Sc. degree in Informatics from Federal University of Para (Brazil), a M.Sc. degree in Informatics from Federal University of Paraiba (Brazil) and a Ph.D. degree in Engineering from École de Technologie Supérieure, University of Quebec (Canada) in 1999, 2002 and 2008, respectively.  Her research interests include pattern recognition, machine learning and computer vision.
\end{IEEEbiography}
\vspace{-30pt}
\begin{IEEEbiography}[{\includegraphics[width=1.0in,height=1.25in,clip]{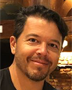}}]
{Alessandro Lameiras Koerich} is an Associate Professor in the Dept. of Software and IT Engineering of the \'{E}cole de Technologie Sup\'{e}rieure (\'{E}TS). He received the B.Eng. degree in electrical engineering from the Federal University of Santa Catarina, Brazil, in 1995, the M.Sc. in electrical engineering from the University of Campinas, Brazil, in 1997, and the Ph.D. in engineering from the \'{E}TS, in 2002. His current research interests include computer vision, machine learning and music information retrieval.
\end{IEEEbiography}
\vspace{-30pt}
\begin{IEEEbiography}[{\includegraphics[width=1in,height=1.25in,clip,keepaspectratio]{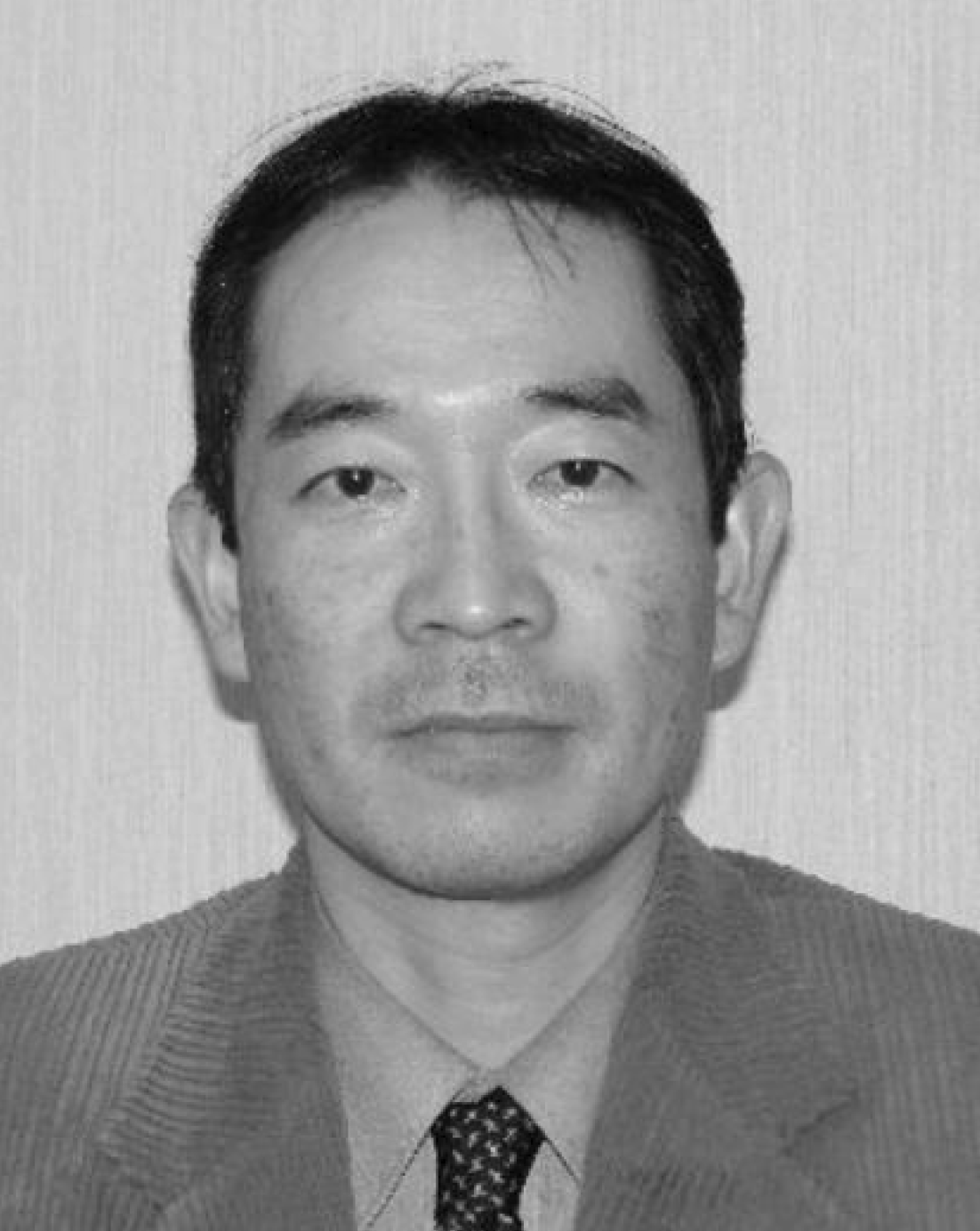}}]{Kazuhiro Fukui}
received his B.E. and M.E. (Mechanical Engineering) from Kyushu University in 1986 and 1988, respectively. In 1988, he joined Toshiba Corporate R\&D Center and served as a senior research scientist at Multimedia Laboratory in 2002. He received his PhD degree from Tokyo Institute of Technology in 2003. He is currently a professor in the Department of Computer Science, Graduate School of Systems and Information Engineering at University of Tsukuba. His interests include the theory of computer vision, pattern recognition, and applications of these theories. He has been serving as a program committee member at many pattern recognition and computer vision conferences, including as an Area Chair of ICPR’12, 14, 16 and 18. He is a member of the IEEE and SIAM.
\end{IEEEbiography}


\vfill



\end{document}


\title{{\it Supplemental material for the paper}\\
Discriminative Singular Spectrum Classifier with Applications on Bioacoustic Signal Recognition}

\author{Bernardo B. Gatto, Juan G. Colonna, Eulanda M. dos Santos, Alessandro L. Koerich and Kazuhiro Fukui
\thanks{B. B. Gatto is with Center for Artificial Intelligence Research (C-AIR),  Tsukuba, Japan e-mail: bernardo@cvlab.cs.tsukuba.ac.jp}%
\thanks{J. G. Colonna is with Institute of Computing, Federal University of Amazonas, Manaus, AM, Brazil e-mail: juancolonna@icomp.ufam.edu.br}%
\thanks{E. M. dos Santos is with Institute of Computing, Federal University of Amazonas, Manaus, AM, Brazil e-mail: emsantos@icomp.ufam.edu.br}%
\thanks{A. L. Koerich is with \'Ecole de Technologie Sup\'erieure (\'ETS), Universit\'e du Qu\'ebec, Montreal, QC, Canada e-mail: alessandro.koerich@etsmtl.ca}
\thanks{K. Fukui is with Center for Artificial Intelligence Research (C-AIR),  Tsukuba, Japan e-mail: kfukui@cs.tsukuba.ac.jp}%
\thanks{Manuscript submitted. March, 2021}
}

\markboth{Submitted to IEEE/ACM Transactions on Audio Speech and Language Processing.}
{Gatto \MakeLowercase{\textit{et al.}}: Discriminative Singular Spectrum Classifier with Applications on Bioacoustic Signal Recognition}

\maketitle


The supplemental material includes additional information about selected materials that were shortly addressed in the paper. First, we give the main notations employed in the paper in Section~\ref{sub:notations}. Next, we present a detailed complexity analysis of MSSA and DSSC and its algorithms in Section~\ref{sub:complexity}. We show the results achieved by E2E systems (1D and 2D-CNNs) on the anuran dataset in Section~\ref{sub:E2E_systems}. Confusion matrices produced by MSSA and DSSC on the anuran dataset are given in Section~\ref{sub:confusion}. Finally, in Section~\ref{sub:spectogram} we show spectrograms of some samples employed in the experiments.

\section{Summary of main notations used in the paper}
\label{sub:notations}

Here we present a comprehensive list of notations employed in the paper (Table~\ref{tab:notations}).

\begin{table}[!h]
\centering
\caption{Summary of main notations used in the paper.}
\footnotesize
\label{tab:notations}
\begin{tabular}{cl}
\toprule
\multicolumn{1}{l}{Notation} & Description                                \\ \midrule

$n$             & number of training samples                              \\
$n_i$           & number of training samples in the $i$-th class          \\
$c$             & number of bioacoustic classes                           \\
$y$             & signal label                                            \\
$X$             & input signal                                            \\
$H$             & Hankel matrix of the signal $X$                         \\
$l$             & maximum time lag of autocorrelation                     \\
$A$             & auto-correlation matrix of $H$                          \\
$U$             & basis vector representing the Hankel matrix $H$         \\
$\mathcal{P}$   & subspace spanned by the selected eigenvectors of $U$    \\
$\mathcal{D}$   & difference subspace \\
$\mathcal{S}$   & sum subspace  \\
$p$             & dimension of the $\mathcal{P}$ subspace                 \\
$d$             & dimension of the $\mathcal{D}$ subspace                 \\
$\phi$, $\psi$  & eigenvectors                                            \\
$\sigma$, $\lambda$, $\delta$  & eigenvalues                              \\ 
\bottomrule
\end{tabular}
\end{table}

\newpage

\section{Algorithm and Computational Complexity Analysis}
\label{sub:complexity}

Except for reducing model parameters, the computational complexity is also an important aspect in the real application of bioacoustic systems. In this section, we calculate the complexity of the training and testing stages of the proposed model. The procedure to perform the bioacoustic classification is as follows. First, Algorithm~\ref{alg:toeplitz} performs a sliding window, producing the Hankel representation of the bioacoustic signals. Next, Algorithm~\ref{alg:subspace} computes the basis vectors of the Hankel matrices, followed the basis selection. In Algorithm~\ref{alg:diff_sub}, a discriminative space is derived. Finally, Algorithm~\ref{alg:DSSC} projects the subspaces produced by Algorithm~\ref{alg:subspace} onto the discriminative subspace followed by a classification based on the nearest subspace.

The complexity of the MSSA is $\mathcal{O}(nl^3)$ in the training phase since one SVD is required for each training sample. Given a testi set with $m$ samples, $\mathcal{O}(ml^3)$ is required to compute the bioacoustic subspaces and $\mathcal{O}(mnl^3)$ to calculate the affinity matrix of the subspaces using the canonical angles, resulting in a complexity of $\mathcal{O}(nml^3)$ in the testing phase.

The complexity of the DSSC is $\mathcal{O}((2n+1)l^3)$ in the training phase since two SVDs are needed for each training sample (one to compute the bioacoustic subspace, and one to obtain its projection onto $\mathcal{D}_{(n)}$). An additional SVD is required to obtain $\mathcal{D}_{(n)}$ from $G_{(n)}$. Given a testing set with $m$ samples, two SVDs are needed for each trial sample and $mn$ ones to compute the affinity matrix, resulting in a complexity of $\mathcal{O}(2ml^3) + \mathcal{O}(mnl^3) = \mathcal{O}(ml^3(2 + n)) = \mathcal{O}(mnl^3)$. This complexity can be further reduced if the number of basis vectors is known in advance. In this case, not all eigenvectors should be estimated. In practical applications, the complexity can be reduced using fast approximate SVD algorithms~\cite{menon2011fast}.

\begin{algorithm}[!htpb]\footnotesize
	\caption{Compute the Toeplitz matrix $H$}\label{alg:toeplitz}\footnotesize
	\textbf{Input:} {$X$, $l$}  \Comment{input signal and its maximum time lag of autocorrelation}\\
	\textbf{Output:} {$H$} \Comment{Toeplitz matrix}
	\begin{algorithmic}[1]
        \State{$H \gets [~]$}
	    \State{$l_X \gets \operatorname{length}(X)$}
        \For{$i \gets 1$ to $l_X-l+1$}
            \State{$X_s \gets X(i:i+l-1)$} \Comment{extract a segment of length $l$ from $X$}
            \State{$H \gets [H~X_{s}^{\top}]$} \Comment{concatenate the segments as columns of $H$, as in Equation (1)} 
        \EndFor
		\State\Return $H$
	\end{algorithmic}
\end{algorithm}

\begin{algorithm}[!htpb]\footnotesize
	\caption{Compute the basis vectors $\overline{U}$ that spans $\mathcal{P}$}\label{alg:subspace}\footnotesize
	\textbf{Input:} {$H$, $p$}  \Comment{input Toeplitz matrix and the subspace dimension}\\
	\textbf{Output:} {$\overline{U}$} 
	\begin{algorithmic}[1]
	    \State{$A \gets HH^{\top}$} \Comment{Equation (2)} 
	    \State{$U \gets \operatorname{svd}(A)$} \Comment{Equation (3)} 
	    \State{$\overline{U} \gets U(1:p)$} \Comment{Equation (4)} 
    	\State\Return $\overline{U}$
	\end{algorithmic}
\end{algorithm}

\begin{algorithm}[!htpb]\footnotesize
	\caption{Compute the basis vectors $B^{\star}$ of the discriminative subspace $\mathcal{D}_{(n)}$}\label{alg:diff_sub}\footnotesize
	\textbf{Input:} {$\{\overline{U}_{i}\}_{i=1}^n$, $n$}  \Comment{set of basis vectors and its cardinality}\\
	\textbf{Output:} {$B^{\star}$} \Comment{basis vectors of the discriminative subspace $\mathcal{D}_{(n)}$}
	\begin{algorithmic}[1]
	
        \State{$G \gets \frac{1}{n}\sum_{i=1}^n \overline{U}_i {\overline{U}_i}^{\top}$} \Comment{Equation (9)} 
        
        \State{$B \gets \operatorname{svd}(G)$}
        \Comment{Equation (10)} 

        \State{$B^{\star} \gets B(d:l)$} \Comment{Equations (13) and (14)} 

		\State\Return $B^{\star}$
	\end{algorithmic}
\end{algorithm}

\begin{algorithm}[!htpb]\footnotesize
	\caption{DSSC}\label{alg:DSSC}\footnotesize
	
	\textbf{Input:} {$\{X_{i}, y_i\}_{i=1}^n$, $X_{inp}$, $n$} \Comment{labeled dataset, its cardinality and an input signal}\\
	\textbf{Output:} {$y$} \Comment{class label of $X_{inp}$}
	\begin{algorithmic}[1]
	
	\State{Compute $\{H_{i}\}_{i=1}^n$ and $H_{inp}$ using Algorithm~\ref{alg:toeplitz}}
	\State{Compute $\{\overline{U}_{i}\}_{i=1}^n$ and $U_{inp}$ using Algorithm~\ref{alg:subspace}}

    \State{Compute $B^{\star}$ using Algorithm~\ref{alg:diff_sub}}

    \For{$i \gets 1$ to $n$}
        \State{$\dot{U}_{i} \gets \operatorname{orth}\left({B^{\star}}^{\top} U_i\right)$}
        \Comment{project the subspaces onto $\mathcal{D}_{(n)}$, as in Equation (12)}
    \EndFor
	
	\State{$\dot{U}_{inp} \gets \operatorname{orth}\left({B^{\star}}^{\top} U_{inp}\right)$}
	\Comment{project the input subspace $\mathcal{P}_{inp}$ onto $\mathcal{D}_{(n)}$, as in Equation (12)}
	
	\State{$S^{\star} \gets 0$} \Comment{highest similarity between $\mathcal{P}_{inp}$ and $\mathcal{P}_{i}$}
	\State{$y^{\star} \gets 0$} \Comment{current label of the corresponding nearest subspace $\mathcal{P}_{i}$}
	\For{$i \gets 1$ to $n$}
        \State{$W_{i} \gets \dot{U}_{i}^{\top}{\dot{U}_{inp}}$}      \Comment{Equation (5)}
        \State{$S_{i} \gets \gamma({\mathcal{P}_1,~\mathcal{P}_2})$}
        \Comment{Equation (8)}
        \If {$S^{\star} < S_i$}
            \State $S^{\star} \gets S_i$
            \State $y^{\star} \gets y_i$
        \EndIf    

    \EndFor

    \State $y \gets y^{\star}$

	\State\Return $y$
	\end{algorithmic}
\end{algorithm}

\section{Results achieved by E2E systems on anuran dataset}
\label{sub:E2E_systems}


Table~\ref{tab:CNN1D} presents the results achieved by two E2E approaches based on 1D and 2D CNN architectures that have been recently used in environmental sound and music genre classification to deal with audio signals of variable length~\cite{len_1,9207309}.
Both E2E approaches split the downsampled audio signal (22.05 kHz) into short segments of 300 ms using a sliding window with $75\%$ of overlapping and filter out segments with low energy, which are likely to be soundless or background noise.
Each segment retains the same label as the original audio samples. While the input of the 1D-CNN is the audio segments, an additional layer is used to convert such segments into short-time Fourier transform spectrograms, which are the input of the 2D-CNN.


In the classification step, since the input audio is split into several segments, we need to aggregate the predictions of the 1D-CNN using a majority vote rule to come up with a final decision on the input audio. The same aggregation is carried out for the predictions of the 2D-CNN~\cite{9207309}.
The low energy filter acts as a feature selection, improving the quality of the training segments employed by both 1D-CNN and 2D-CNN models.

In turn, MSSA and DSSA do not require a data selection mechanism since the dimensionality reduction process provided by the eigenvalues' hierarchical distribution naturally selects the highest energy dimensions, providing an automatic data selection mechanism.
The results achieved by both CNNs are not directly comparable to the results presented in Table III of the paper
due to the differences in the experimental protocol (LOSO CV versus 3-fold CV). However, they show that the proposed approach is very competitive. Besides that, the 1D-CNN has about 620k trainable parameters, and the 2D-CNN has 8.1M trainable parameters, which may prevent their application when limited hardware resources are available.
\begin{table}[]
\centering
\footnotesize
\caption{Results in terms of Precision (Pr), Recall (Re) and F-Score (F-Sc) achieved by E2E 1D-CNN and 2D-CNN on the anuran dataset using 3-fold CV.}
\label{tab:CNN1D}
\begin{tabular}{lcccccc}
\toprule
 \multicolumn{1}{c}{\multirow{2}{*}{Species}}           & \multicolumn{3}{c}{1D-CNN} & \multicolumn{3}{c}{2D-CNN} \\
           \cmidrule(l){2-4}\cmidrule(l){5-7}
        & Pr & Re  &  F-Sc & Pr & Re  &  F-Sc\\
\midrule
\texttt{(a)} \textit{Adenomera andreae} & 0.67 & 0.75 & 0.71 & 0.64 & 0.88 & 0.74 \\
\texttt{(b)} \textit{Ameerega trivittata} & 0.50 & 0.20 & 0.29 & 1.00 & 0.40 & 0.57 \\
\texttt{(c)} \textit{Adenomera hylaeda.} & 0.83 & 0.91 & 0.87 &0.85 & 1.00 & 0.92  \\
\texttt{(d)} \textit{Hyla minuta} & 1.00 & 0.91 & 0.95 & 1.00 & 1.00 & 1.00\\ 
\texttt{(e)} \textit{Hypsiboas cinerasc.} & 0.67 & 1.00 & 0.80 & 0.50 & 0.50 & 0.50 \\
\texttt{(f)} \textit{Hypsiboas cordobae}  & 0.80 & 1.00 & 0.89 & 1.00 & 1.00 & 1.00 \\  
\texttt{(g)} \textit{Leptodactylus fuscus} & 0.40 & 0.50 & 0.44 & 1.00 & 0.25 & 0.40 \\
\texttt{(h)} \textit{Osteocephalus ooph.} & 0.00 & 0.00 & 0.00 & 0.50 & 0.67 & 0.57 \\
\texttt{(i)} \textit{Rhinella granulosa} & 0.60 & 0.60 & 0.60 & 0.83 &1.00 & 0.91 \\
\texttt{(j)} \textit{Scinax ruber} & 0.83 & 1.00 & 0.91 & 1.00 & 0.80 & 0.89  \\ 
\midrule
Micro-accuracy  & -- & -- &  0.75  & -- & -- & 0.82 \\
Macro-average & 0.63 & 0.69 & 0.65 & 0.83 & 0.75 & 0.75 \\
Weighted-average & 0.71 & 0.75  &  0.72 & 0.85 & 0.82 & 0.80  \\
\bottomrule
\end{tabular}
\end{table}


\vspace{60pt}

\section{Confusion matrices produced by MSSA and DSSC on the anuran dataset}
\label{sub:confusion}

\begin{figure}[h!]
\centering
\includegraphics[width=\linewidth]{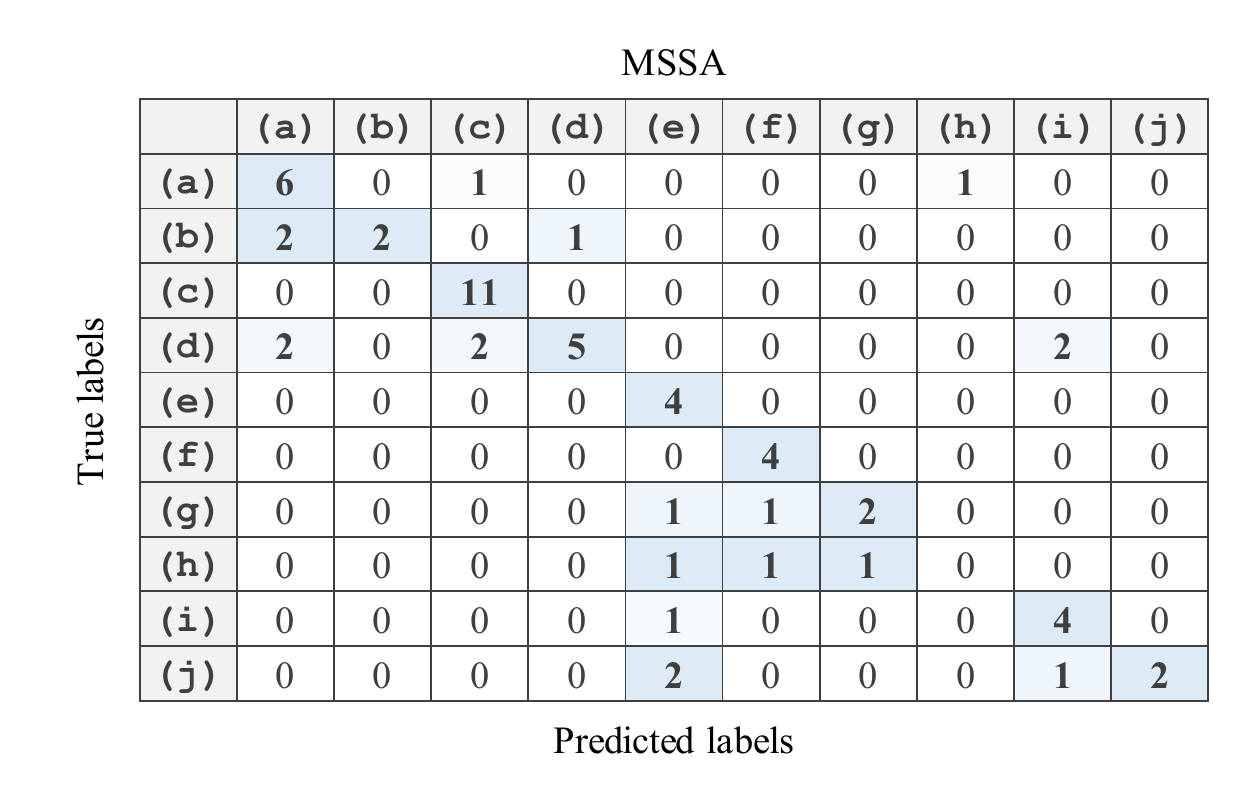}
\caption{Confusion matrix produced by MSSA on the anuran dataset.}
\label{fig:conf_MSSA}
\end{figure}

\begin{figure}[h!]
\centering
\includegraphics[width=\linewidth]{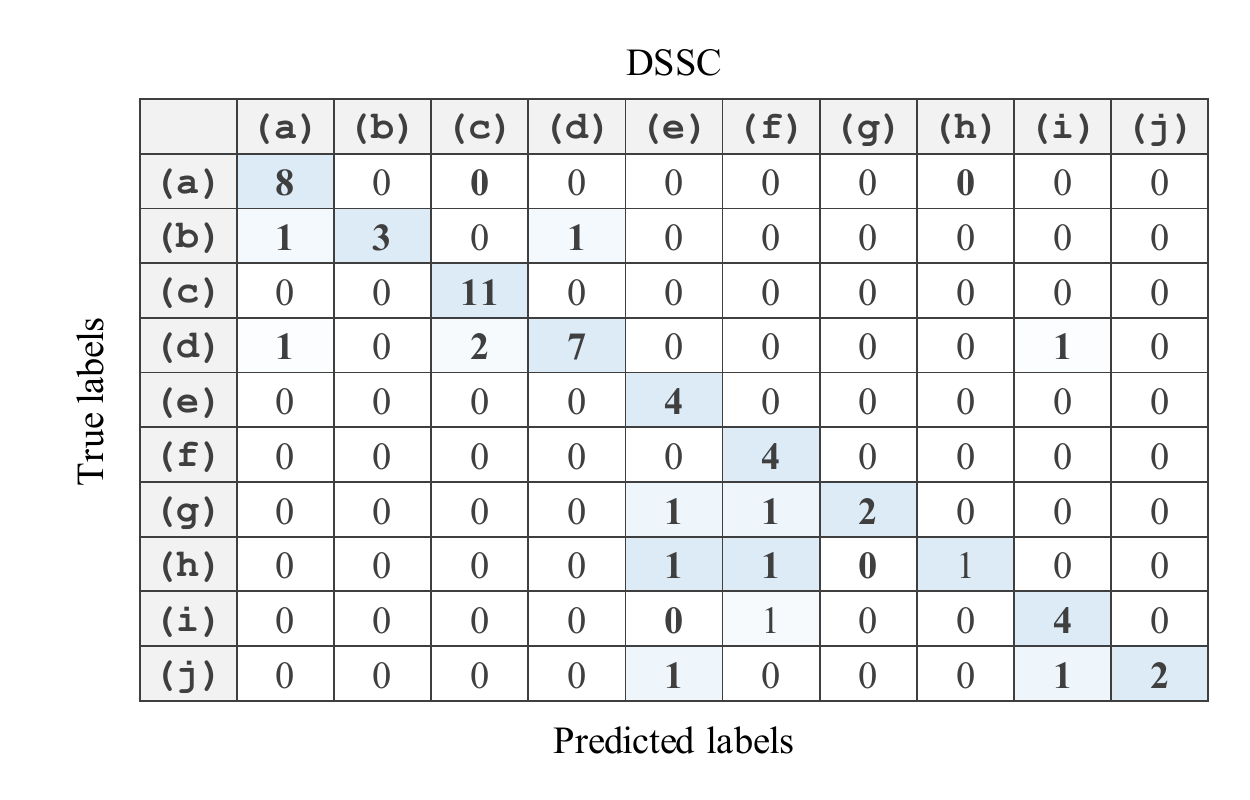}
\caption{Confusion matrix produced by DSSC on the anuran dataset.}
\label{fig:conf_DSSC}
\end{figure}

Figs.~\ref{fig:conf_MSSA} and~\ref{fig:conf_DSSC} show the confusion matrices for the subspace methods. From the confusion matrix, we found that the anuran classes \textit{Hypsiboas cinerascens}, \textit{Osteocephalus oophagus} and \textit{Rhinella granulosa} are often mistakenly classified by MSSA, probably due to the similarity between the basis vectors that represent the common frequencies of these species with others. DSSC improves the classification of the same species but with better discrimination. Motivated by this observation, we consider that the bioacoustic subspaces provided by DSSC can reveal deeper intuitions regarding the main frequencies of bioacoustic signals.

In addition, the confusion matrix of MSSA has shown that the species \textit{Hyla minuta} (row d), was confused with the species \textit{Adenomera andreae} and \textit{Adenomera hylaedactyla}, (columns a and c). The oscillatory components of these species were analysed in Section IV-A of the paper.

\newpage

\newpage

\section{Spectrograms of some samples employed in the experiments}
\label{sub:spectogram}

\begin{figure}[h!]
\centering
\includegraphics[width=0.9\linewidth]{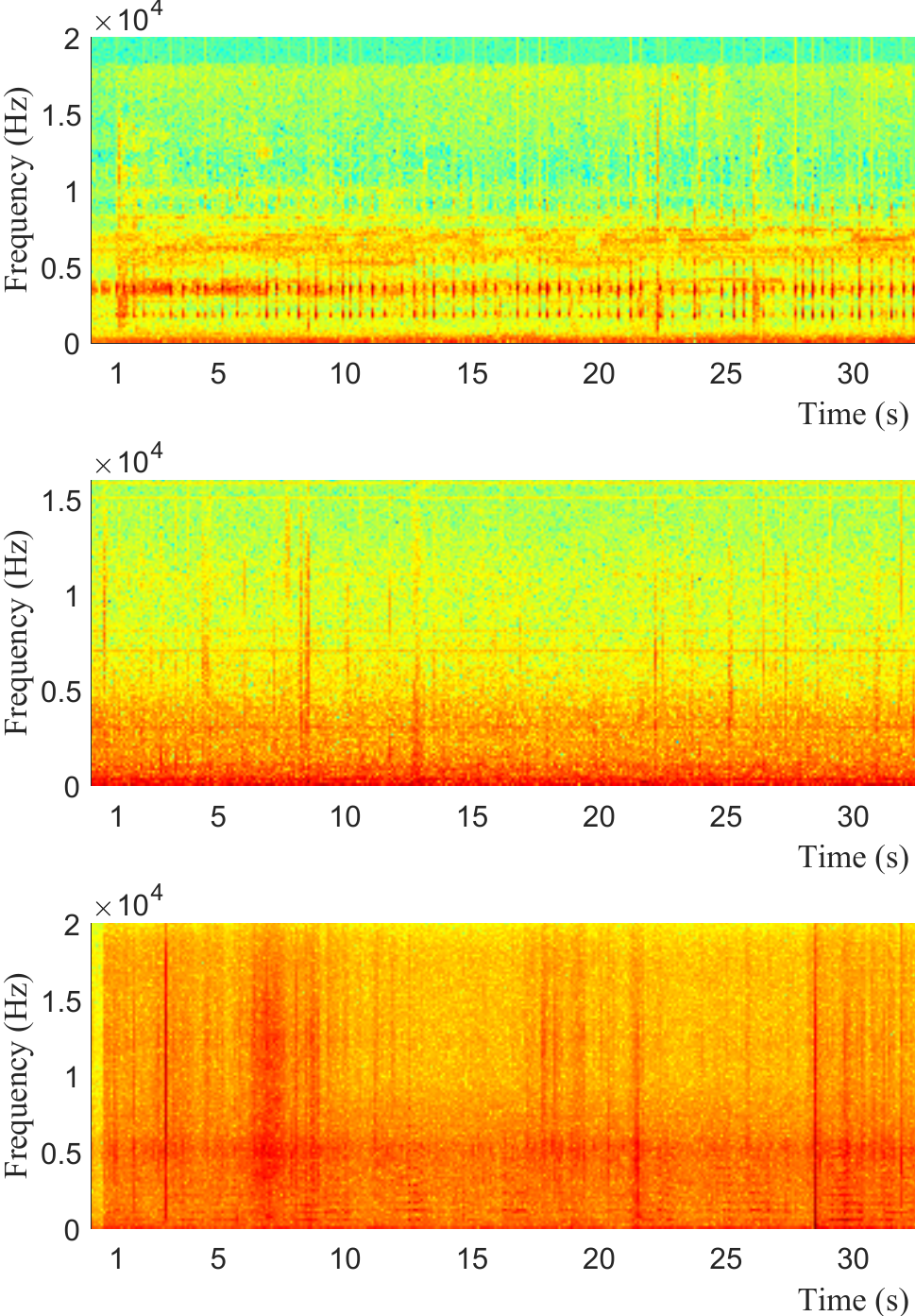}
\caption{The different patterns found in spectrograms of frogs, bees, and mosquitos depict the difficulty of developing bioacoustic recognition systems. The anuran sound (top) presents energy concentrated in the 2 kHz and 4.5 kHz frequency bands with intermittent temporal patterns at regular time stamps. Next, the bee recording (middle) shows overall low-frequency energy. The recording of the mosquito (bottom) shows the spreading of energy in the high-frequency bands. Also, the 6 kHz band is almost continuous, and several temporal spikes reflect the flight pattern of the particular species. Background noises are present in the three spectrograms.}
\label{fig:different_spectogram}
\end{figure}

Fig.~\ref{fig:different_spectogram} shows examples of various acoustic patterns found in these datasets through their spectrograms.

\vfill
\bibliographystyle{IEEEtran}
\bibliography{refs}
